\documentclass[7pt,preprint]{aastex}
\usepackage{multirow}
\shorttitle{New Constraints on Prototypical HyLIRGs}
\shortauthors{Jones et al.}

\begin{document}

\received{}
\revised{}
\accepted{}

\pagenumbering{arabic}

\title {New Constraints on the Molecular Gas in the Prototypical HyLIRGs BRI1202--0725 \& BRI1335--0417}

\author{G. C. Jones\altaffilmark{1,2}, C. L. Carilli\altaffilmark{2,3}, E. Momjian\altaffilmark{2}, J. Wagg\altaffilmark{4}, D. A. Riechers\altaffilmark{5}, F. Walter\altaffilmark{6}, R. Decarli\altaffilmark{6}, K. Ota\altaffilmark{7,3}, R. G. McMahon\altaffilmark{8}}
\altaffiltext{1}{Physics Department, New Mexico Institute of Mining and Technology, Socorro, NM 87801, USA; gcjones@nrao.edu}
\altaffiltext{2}{National Radio Astronomy Observatory, 1003 Lopezville Road, Socorro, NM 87801, USA}
\altaffiltext{3}{Cavendish Astrophysics Group, University of Cambridge, Cambridge, CB3 0HE, UK}
\altaffiltext{4}{Square Kilometre Array Organization, Jodrell Bank Observatory, Lower Withington, Macclesfield, Cheshire SK11 9DL, UK}
\altaffiltext{5}{Department of Astronomy, Cornell University, 220 Space Sciences Building, Ithaca, NY 14853, USA}
\altaffiltext{6}{Max-Planck Institut f\"ur Astronomie, K\"onigstuhl 17, D-69117 Heidelberg, Germany}
\altaffiltext{7}{Kavli Institute for Cosmology, University of Cambridge, Madingley Road, Cambridge CB3 0HA, UK}
\altaffiltext{8}{Institute of Astronomy, University of Cambridge, Cambridge CB3 0HA, UK}

\begin{abstract}
%Intro
We present Karl G Jansky Very Large Array (VLA) observations of CO($J=2\rightarrow1$) line emission and rest-frame 250$\,$GHz continuum emission of the Hyper-Luminous IR Galaxies (HyLIRGs) BRI1202-–0725 ($z=4.69$) and BRI1335–-0417 ($z=4.41$), with an angular resolution as high as 0.15$''$. 
Our low order CO observations delineate the cool molecular gas, the fuel for star formation in the systems, in unprecedented detail. 
%%1202
For BRI1202--0725, line emission is seen from both extreme starburst galaxies: the quasar host and the optically obscured submm galaxy (SMG), in addition to one of the Ly$\alpha$ emitting galaxies in the group. 
%QSO
Line emission from the SMG shows an east-west extension of about $0.6''$.
%Lya2
For Ly$\alpha$-2, the CO emission is detected at the same velocity as [CII] and [NII], indicating a total gas mass $\sim4.0\times10^{10}\,$M$_{\odot}$. 
%%1335
The CO emission from BRI1335--0417 peaks at the nominal quasar position, with a prominent northern extension ($\sim 1''$, a possible tidal feature).  
%Densities
The gas depletion timescales are $\sim 10^7$ years for the three HyLIRGs, consistent with extreme starbursts, while that of Ly$\alpha$-2 may be consistent with main sequence galaxies. 
%Summary
We interpret these sources as major star formation episodes in the formation of massive galaxies and supermassive black holes (SMBHs) via gas rich mergers in the early Universe. 
\end{abstract}


\section{Introduction}
It is well established that massive galaxies form most of their stars
at early times, and the more massive, the earlier (e.g., \citealt{renz06,shap11}).
Hyper-luminous infrared galaxies (HyLIRGs), or galaxies with L$_{IR}(8-1000\mu$m$) >10^{13}$ L$_\odot$ \citep{sand96}
at high redshift discovered in wide field submm surveys,
play an important role in the study of the early
formation of massive galaxies, corresponding to perhaps the dominant
star formation episode in the formation of massive elliptical galaxies
(e.g., \citealt{case14}). Star formation rates over 1000 M$_\odot$
year$^{-1}$, occurring for timescales up to 10$^8$ years, can form the
majority of stars in a large elliptical (e.g., \citealt{nara15}). These
systems are typically highly dust-obscured, and best studied at IR
through radio wavelengths.

A second important finding in galaxy evolution is the correlation
between the masses of supermassive black holes (SMBHs) and their host spheroidal
galaxies (e.g., \citealt{korm13}).  While the exact nature of this correlation
remains under investigation, including its redshift evolution (e.g., \citealt{walt04,wang13,will15,kimb15}),    
and certainly counter examples exist (e.g., \citealt{vand12}), the general
implication of such a correlation would be that there exists some 
sort of co-evolution of supermassive black holes and their host galaxies

Submm galaxies (SMGs) and other ultra-luminous infrared galaxies (ULIRGs), powered by either star 
formation or active galactic nucleus (AGN), have
clustering properties that imply they reside in the densest cosmic
environments (proto-clusters) in the early Universe (e.g., \citealt{blai02,chap09,capa11}).
The mechanism driving the extreme star
formation rates remains uncertain, or at least multivariate.  Some
systems are clearly in the process of a major gas rich merger, in
which nuclear starbursts are triggered by tidal torques driving gas to
the galaxy centers (e.g., \citealt{enge10,tacc08,riec11L31}). However, some
SMGs show clear evidence for smoothly rotating disk galaxies, with
little indication of a major disturbance (e.g., \citealt{hodg12,kari16}).  
There is some evidence that the most extreme luminosity systems,
in particular powerful AGN in HyLIRGs (quasars and powerful radio
galaxies), are preferentially involved in active, gas rich major
mergers leading to compact, nuclear starbursts (e.g., \citealt{riec08,riec11L32,mile08,ivis12}).
These merging systems may indicate a major accretion event in the
formation of the SMBH, coeval with the major star formation episode of
the host galaxy.

Wide field surveys have identified thousands of extreme starbursts in
the early Universe, and the statistical properties and demographics
are reasonably well determined (e.g., \citealt{case12}). Study of such systems are
now turning to the detailed physical processes driving extreme
starbursts in the early Universe, enabled by the advent of sensitive,
wide-band interferometers such as the Atacama Large Millimeter/submillimeter Array (ALMA), 
the Very Large Array (VLA), and the NOrthern Millimeter Extended Array (NOEMA). These
facilities allow for deep, very high resolution imaging of the dust,
gas, star formation, and dynamics in extreme starbursts, unhindered by
dust obscuration.  Key questions can now be addressed, such as: What
is the relationship between gas mass and star formation (i.e., the `star formation
law')? What are the interstellar medium (ISM) physical conditions that drive 
the extreme star formation? What dominates the gas supply?  What role
does the local environment play (proto-cluster, group harassment)?
What is the role of feedback in mediating galaxy formation, driven by
either AGN and/or starbursts?                                                             

The BRI1202--0725 ($z\sim$4.7) and BRI1335--0417 ($z\sim$4.4) systems were
among the first HyLIRG systems discovered at very high redshift
(\citealt{irwi91,mcma94,omonA96}), and they 
remain two of the brightest unlensed submm sources known at $z > 4$. 
These two systems are the archetypes for coeval extreme starbursts
and luminous AGN within 1.4 Gyr of the Big Bang. We have undertaken
an extensive study of these systems, using ALMA, the VLA, NOEMA, and
other telescopes, to determine the dominant physical processes driving 
the extreme starbursts and their evolution. 

In this paper, we present our latest VLA observations of the CO($J=2\rightarrow1$)
emission from these two systems, at a resolution as high as $0.15''=1\,$kpc. Imaging of low order CO emission is crucial to understand the
distribution and dynamics of the cool molecular gas fueling star
formation in the systems. We will assume ($\Omega_{\Lambda}$,$\Omega_m$,h)=(0.682,0.308,0.678) \citep{plan15} throughout. At this distance, 1 arcsecond corresponds to 6.63$\,$kpc at $z$=4.69 (BRI1202--0725) and 6.82$\,$kpc at $z$=4.41 (BRI1335--0417).

\section{The Sources}

\subsection{BRI1202--0725}
\begin{figure}[h]
\centering
\includegraphics[scale=0.6,clip=true]{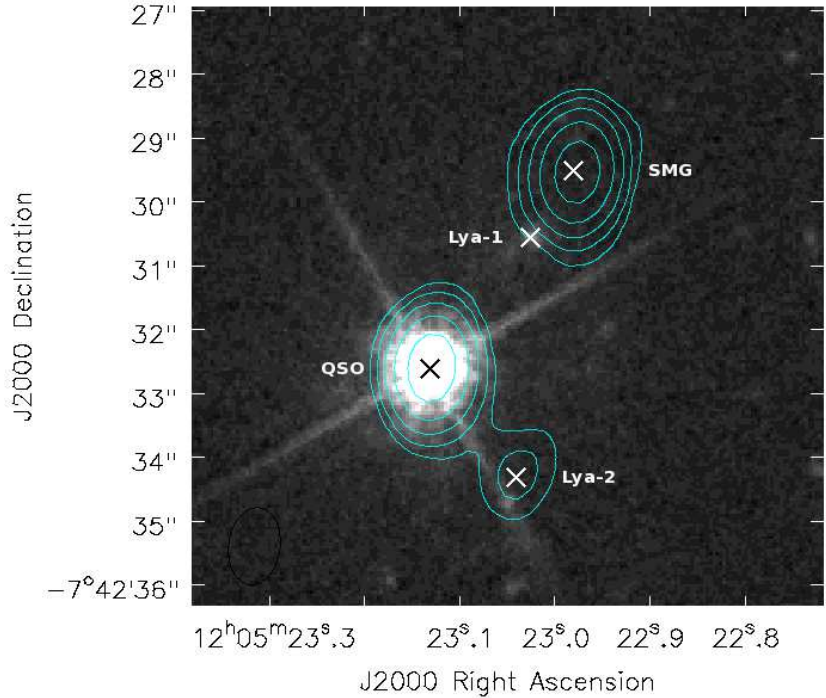}
\caption{The BRI1202--0725 field. Contours show 340$\,$GHz continuum from \citet{cari13}. Contours are a geometric progression of 2, starting at $\pm4\times0.1\,$mJy beam$^{-1}$, and the beam is $1.2''\times0.8''$ at PA -7$^{\circ}$. The background shows the \textit{HST}/ACS F775W filter image of \citet{deca14}. Positions of each object from 340$\,$GHz are marked, expect for Ly$\alpha$-1, whose optical position is marked.}
\label{finder}
\end{figure}

BRI1202--0725 ($z=4.69$) is perhaps the richest physical laboratory to study major gas rich mergers in the early Universe. The system includes two HyLIRGs separated by $4''$: a prototypical, highly dust obscured SMG, and a luminous optical quasar (QSO) in an extreme starburst galaxy. We will refer to the QSO host galaxy as the QSO in what follows. Two Ly-$\alpha$ emitting galaxies (LAEs) are also seen in the system (\citealt{omonA96,hu96}). The 340$\,$GHz emission of \citet{cari13} and the \textit{HST}/ACS F775W filter image of \citet{deca14} are shown in an overview of the field (Fig. \ref{finder}).

\subsubsection{SMG \& QSO}
The first detected component of the system, the QSO, was originally discovered as the highest redshift source in a color-selected (UKST $\mathrm{B_J}$, R, and I plates) survey for high redshift QSOs \citep{irwi91} using the Automated Plate Measuring (APM) machine. This discovery in the optical was followed with a detection of FIR emission from thermal dust \citep{mcma94} with a $12''$ beam. Optical spectra show that it features a central black hole with mass $\sim10^9\,$M$_{\odot}$, as estimated from the MgII linewidth  \citep{carn13}.

The SMG companion was first discovered in the 1.25$\,$mm observations of \citet{omonN96}. These $\sim2''$ resolution images revealed mm continuum and CO($J=5\rightarrow4$) emission in the SMG and QSO, showing that they are a pair of distinct galaxies \citep{cari02}. Spectral energy distribution (SED) fits of \citet{iono06} yield FIR luminosities of 1.2$\times10^{13}\,$L$_{\odot}$ and 3.8$\times10^{13}\,$L$_{\odot}$ for the SMG and QSO, respectively, indicating that both galaxies are HyLIRGS independently.

The field has been well studied in the submm (\citealt{isaa94,benf99,wagg12,cari13}), mm (\citealt{mcma94,omonA96}), and radio continuum (\citealt{yun00,cari02,momj05,wagg14}). The SEDs of these observations are well fit by combined synchrotron, free-free, and dust emission models for star forming galaxies. The dust masses are M$_D\sim10^9\,$M$_{\odot}$ (\citealt{mcma94,isaa94,hu96,benf99}) and the dust temperatures in each object are T$_D\sim 70\,$K (\citealt{isaa94,leec01,wagg14}). Very Long Baseline Interferometry (VLBI) radio continuum observations yield a non-thermal brightness temperature consistent with a nuclear starburst in the QSO and the SMG \citep{momj05}.

CO observations have successfully detected $J=7\rightarrow6$ (\citealt{omonN96,salo12}), $J=5\rightarrow4$ (\citealt{omonN96,ohta96,yun99,salo12}), $J=4\rightarrow3$ \citep{omonA96}, $J=2\rightarrow1$ (\citealt{kawa99,cari02,wagg14}), and $J=1\rightarrow0$ (\citealt{riec06}). The QSO CO emission has a smaller velocity width than the SMG. The different CO transitions show relative line strengths that imply high excitation, essentially thermal excitation, to $J=4\rightarrow3$ \citep{riec08}, $J=5\rightarrow4$ \citep{wagg14}, or up to $J=8\rightarrow7$ \citep{salo12}. 

Assuming a CO luminosity to gas mass conversion factor typical of ULIRGs ($\alpha_{CO}=0.8\,$M$_{\odot}$K$^{-1}$ km$^{-1}$ s pc$^{-2}$; \citealt{bola13}), the molecular gas masses are M$_{gas}\sim10^{11}\,$M$_{\odot}$ (\citealt{mcma94,ohta96,kawa99,cari02,salo12,carn13}). In particular, \citet{wagg14} find gas masses of $(7.0\pm0.6)\times10^{10}\,$M$_{\odot}$ for the SMG and $(4.8\pm0.4)\times10^{10}\,$M$_{\odot}$ for the QSO. The SED fits of \citet{wagg14} give star formation rates (SFRs) of $(4\pm2)\times10^3\,$M$_{\odot}$ year$^{-1}$ for the SMG and for the QSO. These values give short gas depletion timescales of $18\pm9\,$Myr for the SMG and $12\pm6\,$Myr for the QSO and placing them at the low-timescale end, although within the scatter, of values seen in extreme starbursts in the early Universe (\citealt{hodg15,kenn12}).

Imaging has been performed in the atomic fine structure lines, including  [CI] \citep{salo12}, [NII] \citep{deca14}, and [CII] (\citealt{iono06,wagg12,cari13,carn13}). The [CII] imaging in particular shows an ordered velocity gradient across the SMG, suggesting rotation on scales of a few kpc \citep{cari13}. If this is the case, the dynamical mass is M$_{dyn}\sim 4\times 10^{11}\,$M$_{\odot}$. The velocity field of the QSO is less ordered, but [CII] observations suggest a small-scale outflow \citep{wagg12,cari13,carn13}.

\subsubsection{Ly-$\alpha$ Emitters}
The Ly-$\alpha$ emission in the BRI1202--0725 field shows a broad line from the QSO, and two additional sources: a detected extension 2.3$''$ to the north of the QSO \citep{peti96,hu96,will14}, called Ly$\alpha$-1, and emission from a galaxy $\sim3''$ to the southwest of the QSO, called Ly$\alpha$-2 \citep{hu97,will14}. Optical continuum emission from both Ly$\alpha$-1 (\citealt{font96,hu96,font98}) and Ly$\alpha$-2 (\citealt{will14}) is also detected. Ly$\alpha$-1 shows a rest frame \textit{B} magnitude comparable to a L* galaxy (\citealt{hu96}). 

A study of the permitted UV atomic emission lines and the forbidden FIR fine structure lines from Ly$\alpha$-1 suggest that the QSO is unlikely to be the heating source for the ionized gas, and the line ratios are consistent with heating by local star formation \citep{will14}. Ly$\alpha$-1 may then be shielded by an obscuring torus around the QSO. This face-on geometry relative to us is consistent with the fact that we observe the broad line region of the QSO in strong UV lines, even though the host galaxy has a large dust mass. 

The [CII] emission shows Ly$\alpha$-1 may be in a complete bridge of atomic gas between the QSO and SMG \citep{cari13}. Such a bridge would be the natural consequence of strong gravitational interaction during the merging of two gas rich galaxies, although the detection of the bridge remains tentative. Using a stellar population synthesis model, \citet{font96} found a stellar mass of M$_{stellar}\sim10^9\,$M$_{\odot}$. The SFR of Ly$\alpha$-1 ($\sim15\,$M$_{\odot}$ year$^{-1}$) was originally estimated using a similar synthesis model \citep{font98}, while \citet{cari13} found $\sim19\,$M$_{\odot}$ year$^{-1}$ using a $\mathrm{L_{[CII]}}$ conversion. 

Evidence suggests that Ly$\alpha$-2 may be somehow associated with an outflow from the QSO. In particular, an extension from the QSO towards Ly$\alpha$-2 has been seen in CO($J=5\rightarrow4$) \citep{salo12}, and emission between the QSO and Ly$\alpha$-2 appears in the [CII] channel maps of \citet{cari13}. The star formation rate, estimated from the dust continuum emission, is $\sim170\,$M$_{\odot}$ year$^{-1}$ \citep{cari13}.

Ly-$\alpha$ emission from Ly$\alpha$-1 and Ly$\alpha$-2 shows full width at half maximum (FWHM) $\sim$1400 and 1200$\,$km s$^{-1}$ \citep{will14}, which are much broader than their [CII] emission (FWHM$\sim$50 and 340 km s$^{-1}$; \citealt{cari13}).

While \citet{ohta00} detected [OII] in Ly$\alpha$-1 and \citet{deca14} found [NII] in Ly$\alpha$-2, deep Very Large Telescope (VLT) observations did not detect UV lines in either source (no NV, SiIV, CIV, HeII; \citealt{will14}). These non-detections suggest that the LAEs originate from star formation, not QSO photoionization, and trace a turbulent ISM/star formation environment affected by the QSO outflow.

%From a simple virial calculation, the two HyLIRGs and Ly$\alpha$-1 appear to be a gravitationally bound cluster, while Ly$\alpha$-2 may be an escaping outburst. Similarly, treating the four galaxies as a point mass at the position of their center of mass results in three sources having v$<$v$_{esc}$, but Ly$\alpha$-2 showing v$\sim2$v$_{esc}$. Adding the contribution of a dark matter halo of $10^{11}\,$M$_{\odot}\sim0.5\,$M$_{gas}$ has little effect, while a halo mass of $>10^{12}\,$M$_{\odot}$ causes the escape velocity of Ly$\alpha$-2 to exceed its observed velocity. These dark matter halo masses are comparable with those estimated by \citet{mand06} ($10^{11-13}\,$M$_{\odot}$). Future high resolution data will enable us to determine the velocity field of the sources more exactly, making more precise cluster dynamics possible.

\subsection{BRI1335--0417 ($z$=4.4)}
Also discovered in the APM survey \citep{irwi91}, BRI1335--0417 ($z$=4.4)  is likely a late-stage, gas rich, ``wet'' merger, with one highly disturbed galaxy associated with an AGN \citep{riec08}. The molecular gas shows a dominant southern core, associated with the QSO host galaxy, and a $\sim1''$ extension towards a weaker northern peak \citep{cari02,riec08}. It is also classified as a HyLIRG due to its FIR luminosity of 3.1$\times10^{13}$L$_{\odot}$ \citep{cari02}. The relation of $5100\,$\AA$\,$ luminosity and H$\beta$ linewidth to black hole mass predicts M$_{BH}\sim10^{10}\,$M$_{\odot}$ \citep{shie06}. Similarly to BRI1202--0725, the rest frame radio through FIR SED of the system is consistent with a star forming galaxy \citep{wagg14}. SED modeling suggests a total dust mass of M$_D\sim10^{8-9}M_{\odot}$ (\citealt{omonA96,benf99,riec08}) and a dust temperature T$_D\sim50$K (\citealt{benf99,cari99}).

CO observations have revealed both $J=2\rightarrow1$ (\citealt{cari99,cari02,riec08,wagg14}) and $J=5\rightarrow4$ \citep{guil97}. Again, the CO spectral line energy distribution (SLED) implies very high excitation and thermalized line strengths up to $J=5\rightarrow4$ (\citealt{wagg14,guil97}). The standard ULIRG L$'_{CO}$ to molecular gas mass conversion factor gives a molecular gas mass of $\sim10^{11}$M$_{\odot}$ (\citealt{guil97,cari99,cari02,riec08,wagg14}).

[CII] emission was detected in this source with the Atacama Pathfinder EXperiment telescope (APEX) \citep{wagg10}. 

The SED fits of \citet{wagg14} give an SFR of $(5\pm1)\times10^3\,$M$_{\odot}$ year$^{-1}$. When combined with their gas mass estimate of $(5.8\pm0.5)\times10^{10}\,$M$_{\odot}$, they find a low gas depletion timescale of $12\pm3\,$Myr, placing this source in the starbursting area of a Kennicutt-Schmidt diagram.

VLBI radio continuum observations did not detect any compact radio source, consistent with the idea that the radio continuum emission is due to star formation, not a compact AGN \citep{momj07}.

\section{Observations}

Observations were carried out between September 2013 and February 2014
with the VLA of the NRAO
in the B-- and BnA--configurations, as well as during the move time between the two configurations. Each observing
session was 2 hours, consisting of time on the target
(BRI1202--0725/BRI1335--0417), complex gain calibrator
(J1229+0203/J1338--0432), and 3C286, which was used for bandpass, delay, and flux calibration. Table \ref{obstime} summarizes the observing sessions for each
target.

\begin{deluxetable}{lccc}
\tablecolumns{4}
\tablewidth{0pt}
\tablecaption{VLA Observations \label{obstime}}
\tablehead{ \colhead{Source} & \colhead{Date} & \colhead{Configuration} & \colhead{Time [hr]}}
\startdata
 BRI1202--0725&2013 Oct 28 & B&2 \\
 BRI1202--0725&2014 Jan 23 & B$\rightarrow$BnA&2\\
 BRI1202--0725&2014 Jan 30 & BnA&2 \\
 BRI1202--0725&2014 Feb 5 & BnA &2\\
 BRI1202--0725&2014 Feb 9 & BnA &2\\
 BRI1202--0725&2014 Feb 10 & BnA&2 \\ \hline
 
 BRI1335--0417&2013 Sep 29 & B&$2\times2$ \\
 BRI1335--0417&2014 Jan 18 & B&2 \\
 BRI1335--0417&2014 Jan 19 & B&$2\times2$ \\
 BRI1335--0417&2014 Jan 21 & B&$2\times2$ \\
 BRI1335--0417&2014 Jan 23 & B$\rightarrow$BnA&$2\times2$ \\
 BRI1335--0417&2014 Jan 24 & B$\rightarrow$BnA&2 \\
 BRI1335--0417&2014 Feb 5 & BnA&2 \\
 BRI1335--0417&2014 Feb 8 & BnA&$2\times2$ \\
 BRI1335--0417&2014 Feb 9 & BnA&2 \\
 BRI1335--0417&2014 Feb 10 & BnA&2 \\\hline 
\enddata
\end{deluxetable}

In the following, we define zero velocity of the CO($J=2\rightarrow1$)
line as its frequency at CO redshifts of \citet{wagg14} for the
BRI1202--0725 SMG ($z=4.6915$), BRI1202--0725 QSO ($z=4.6942$), and
BRI1335--0417 ($z=4.4065$). This shifts the CO($J=2\rightarrow1$) line
($\nu_{rest}=230.538$ $\,$GHz) to observed frequency of $\nu_{obs}=40.5057\,$GHz,
40.4865$\,$GHz, and 42.6409$\,$GHz, respectively.

We used the Q-band receiver and the four pairs of the 3-bit samplers on each VLA antenna to deliver a total of 8$\,$GHz ($4\times2\,$GHz; $40\,$GHz to $48\,$GHz), with each $2\,$GHz being in right and left circular polarization. During correlation, each $2\,$GHz was further split into 16 spectral windows (SPWs) of width $128\,$MHz, and each SPW was split to 64 $2\,$MHz wide channels. To correct for the poor sensitivity of the SPWs at their edges, two tunings were observed with a 16$\,$MHz offset (39.960 to 48.008$\,$GHz and 39.976 to 48.024$\,$GHz) to mitigate the ranges of poor sensitivity in the instrument's frequency response.

Data were calibrated using the Common Astronomy
Software Applications (CASA) and altered versions of its version 4.1.0 calibration pipeline. First, the
data were processed by a pipeline that was optimized for spectral line
data. This optimization entailed the removal of Hanning smoothing and the flagging of only the one end channel 
of each SPW, rather than the end three channels. In addition, the two (B$\rightarrow$BnA) or three (BnA) 
northernmost antennas were flagged to create a circular beam.
The data were then inspected for troublesome antennas or
frequency ranges, which were manually flagged. The flagged measurement
set was then passed through a final, calibration-only pipeline that did not flag any additional data. The
six measurement sets of BRI1202--0725 and fifteen measurement sets of
BRI1335--0417 were used to create naturally weighted
images. Additionally, D--configuration VLA data of both
objects from \citet{wagg14} was used. The absolute flux scale uncertainty was $\sim3\%$ \citep{perl13}. Analysis was performed with both
Astronomical Image Processing System (AIPS) and CASA.

Images were created with the CASA task \textit{clean}, using natural
weighting. Continuum images were made by combining the data of all
observation sessions using multi-frequency synthesis (MFS), excluding
channels that contained line emission. Spectral line cubes were made
by applying radial \textit{uv}-tapers to the B--configuration data,
creating images with different synthesized beam sizes to investigate
limits to the CO source sizes. We also explored combining the
D-- and B--configuration \textit{uv}-data before imaging. However, the synthesized beam
solid angles differed by two orders of magnitude, and the resulting
combined synthesized beam had very broad wings. We explored
extensively different visibility weighting schemes in order to restore
jointly any large and small scale structure, while retaining adequate
sensitivity. Unfortunately, for faint sources such as these, we found
that it was difficult to differentiate conclusively between real
extended structure and apparent structure caused by the broad wings of
the combined synthesized beam. Hence, we have taken a conservative
approach to determining source sizes by comparing D--configuration spectra with
B--configuration spectra generated using gradual tapers of the visibilities.

\section{Results}

\subsection{Continuum}
We first consider the 44$\,$GHz continuum emission. 
In both sources, the SED fitting of broad band continuum emission implies 
that the 44$\,$GHz observed emission (250$\,$GHz rest frame), is thermal emission 
from cold dust \citep{wagg14}.

\begin{figure}[h]
\centering
\includegraphics[scale=0.4,clip=true]{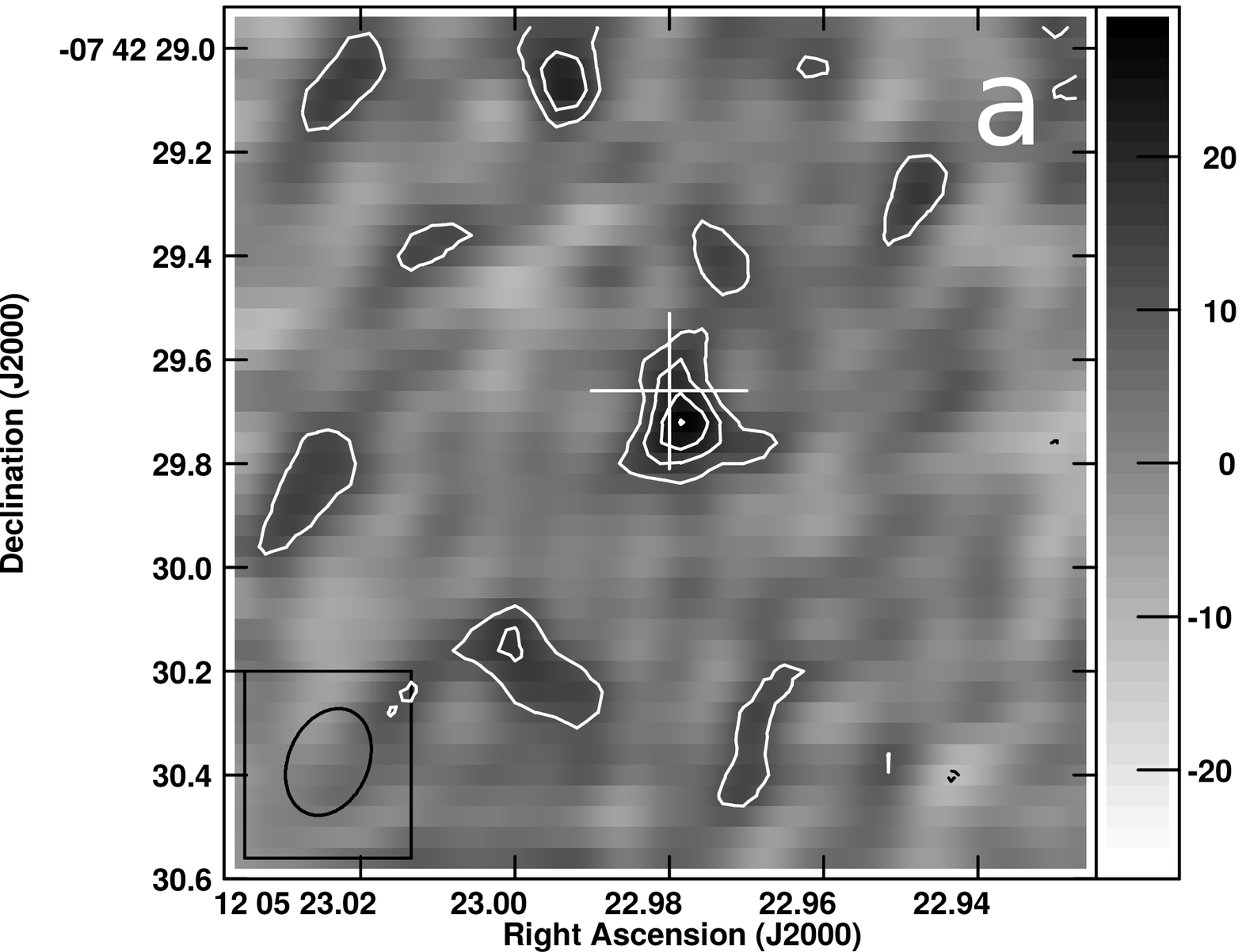}
\includegraphics[scale=0.4,clip=true]{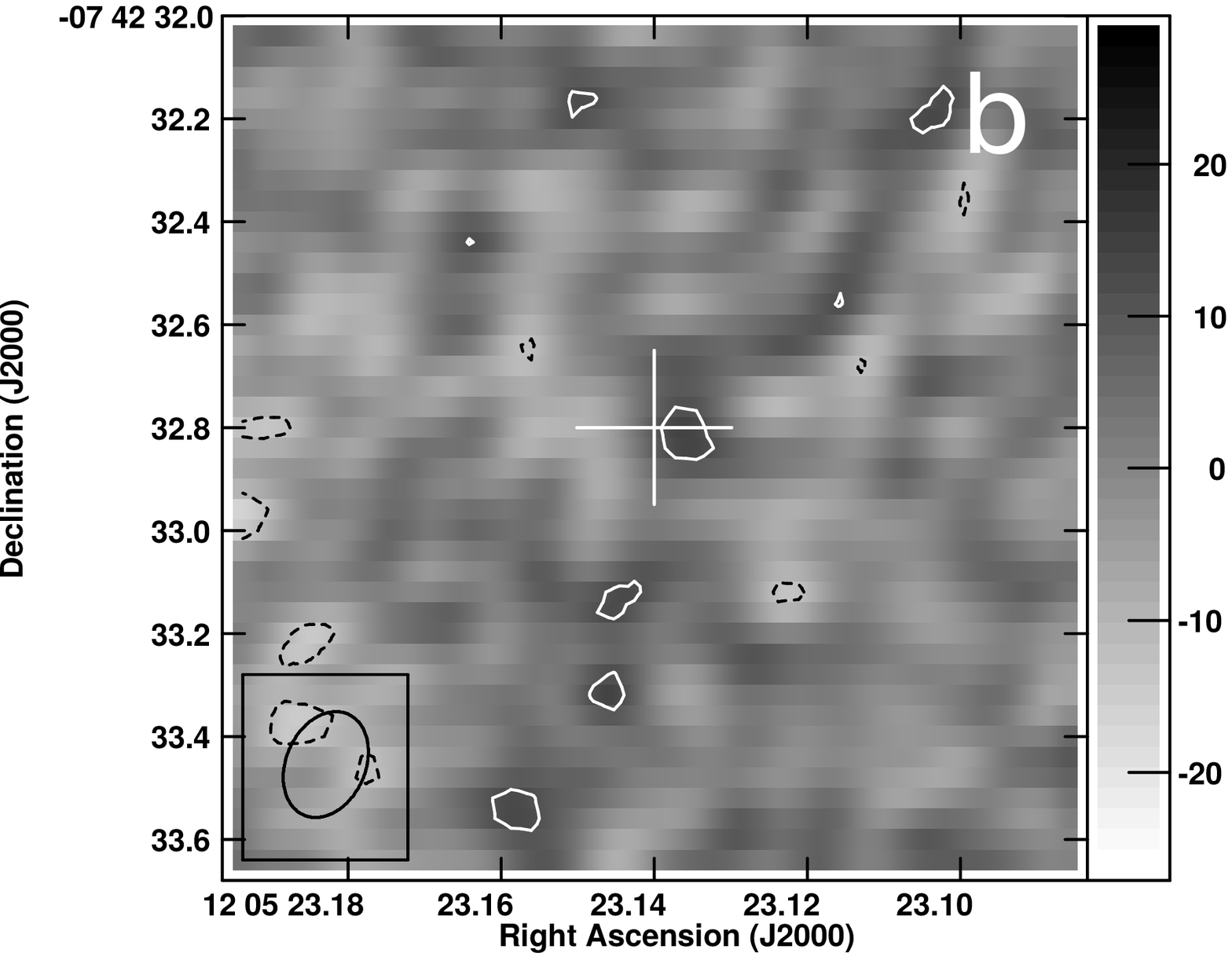}
\caption{Naturally weighted VLA B--configuration 44$\,$GHz continuum images of the SMG (a) and QSO (b) in BRI1202--0725. Contours begin at $\pm2\sigma$, where $1\sigma= 5.6\,\mu$Jy beam$^{-1}$, and are in steps of $1\sigma$. Crosses represent source locations from CO($J=2\rightarrow1$) emission. Restoring beam $(0.21''\times0.15'')$ at position angle $-24^{\circ}$ shown as an ellipse in the bottom left corner. The greyscale is in units of $\mu$Jy beam$^{-1}$.}
\label{cont1}
\end{figure}

Fig. \ref{cont1}a shows the VLA B--configuration
44GHz continuum image of the SMG in BRI1202--0725, at a
resolution of 0.21$''\times0.15''$ at PA = -24$^{\circ}$. A two--dimensional Gaussian fit
to the emission seen in this image yields a size of (0.18$'' \pm0.05'')\times(0.14''\pm0.08''$) at PA = 2$^{\circ}$, or $(1.2\pm0.3)\times (0.9\pm 0.5)\,$kpc$^2$. 
The peak and integrated flux densities of the SMG fit are
$23\pm6\,\mu$Jy beam$^{-1}$ and $41\pm6\,\mu$Jy, respectively. The D--configuration 
analysis at $2''$ by \citet{wagg14} found $51\pm 6\,\mu$Jy. 
These nearly equal integrated flux densities suggest that the majority of the continuum emission originates from the central $1\,$kpc, not from diffuse structures. 

For the QSO host galaxy in 1202--0725, the continuum emission is weak (see Fig. \ref{cont1}b),
and we can only determine an upper limit to the source size of
$<(1.4 \times 1.0)\,$kpc$^2$. From Gaussian fitting, the peak and
integrated flux densities of the QSO are $14\pm6\,\mu$Jy beam$^{-1}$
and $16\pm5\,\mu$Jy. \citet{wagg14} D--configuration analysis found $24\pm
6\,\mu$Jy. 

\citet{momj05} observed the synchrotron emission of BRI1202--0725 using VLBI at 
1.4$\,$GHz. The SMG was clearly resolved, and larger than, but comparable to
the sizes observed in the 44GHz dust continuum ($1.91\times1.10\,$kpc$^2$).

\begin{figure}[h]
\centering
\includegraphics[scale=0.4,clip=true]{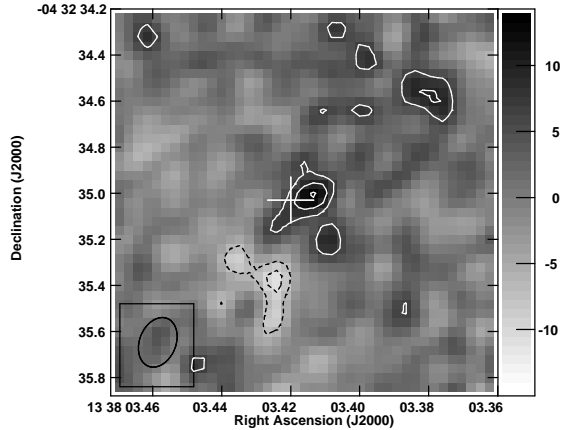}
\caption{Naturally weighted VLA B--configuration 44$\,$GHz continuum image of BRI1335--0417. Contours begin at $\pm2\sigma$, where $1\sigma= 3.1\,\mu$Jy beam$^{-1}$, and are in steps of $1\sigma$. Cross represents source location from CO($J=2\rightarrow1$). Restoring beam $(0.22''\times0.16'')$ at position angle $-22^{\circ}$ shown as an ellipse in the bottom left corner. The greyscale is in units of $\mu$Jy beam$^{-1}$.}
\label{cont2}
\end{figure}

BRI1335--0417 is detected in continuum emission with a peak at the 4$\sigma$
level (Fig. \ref{cont2}). A two--dimensional Gaussian fit
to the source yields a size of $(0.28''\pm0.06'')\times(0.2''\pm0.1'')$ at PA = 15$^{\circ}$, or $(1.9\pm 0.4)\times(1.2\pm0.7)\,$kpc$^2$, a peak flux density of $10\pm3\,\mu$Jy beam$^{-1}$,
and an integrated flux density of $24\pm 5\,\mu$Jy. \citet{wagg14}
reported a D--configuration flux density of $40\pm7\,\mu$Jy, which is
$\sim3\sigma$ greater than our value. A VLBI observation of BRI1335--0417 at 1.4$\,$GHz yielded a size of $1.29\times0.77\,$kpc$^2$ \citep{momj07}.

We do not detect BRI1202--0725 Ly$\alpha$-2 at 44$\,$GHz, implying a lower size limit of $>0.2''$, or $>1.3\,$kpc.

\subsection{Peak Spectra Size Constraints}

In order to get a firm handle on the CO source size limits, we compare
B--configuration CO spectra obtained using different \textit{uv}--tapers, corresponding
to different spatial resolutions, to much lower spatial resolution D--configuration spectra ($\sim 2''$). In each case, the B--configuration spectrum was
extracted at the peak of the velocity integrated emission (moment zero image) for the given
resolution.  

We applied circular Gaussian \textit{uv}--tapers to our B--configuration data
in units of kilo--wavelengths ($\lambda\sim 7.5\,$mm for BRI1202--0725
\& $\lambda \sim 7.1\,$mm for BRI1335--0417). We included tapers of
FWHM 750k$\lambda$ (hereafter B750), 450k$\lambda$ (B450), and
250k$\lambda$ (B250). Note that while these tapers increased the
effective beam size of the configuration, these are still much
smaller than the $2''$ resolution of the D--configuration whose maximum baseline corresponds to $\sim140\,$k$\lambda$. The beam
sizes of the tapered and untapered data are listed in Table
\ref{north1}. Note that for tapers lower than 250k$\lambda$, the noise
increases dramatically due to the lack of short spacings. 

\begin{figure}[h]
\centering
\includegraphics[scale=0.5,clip=true]{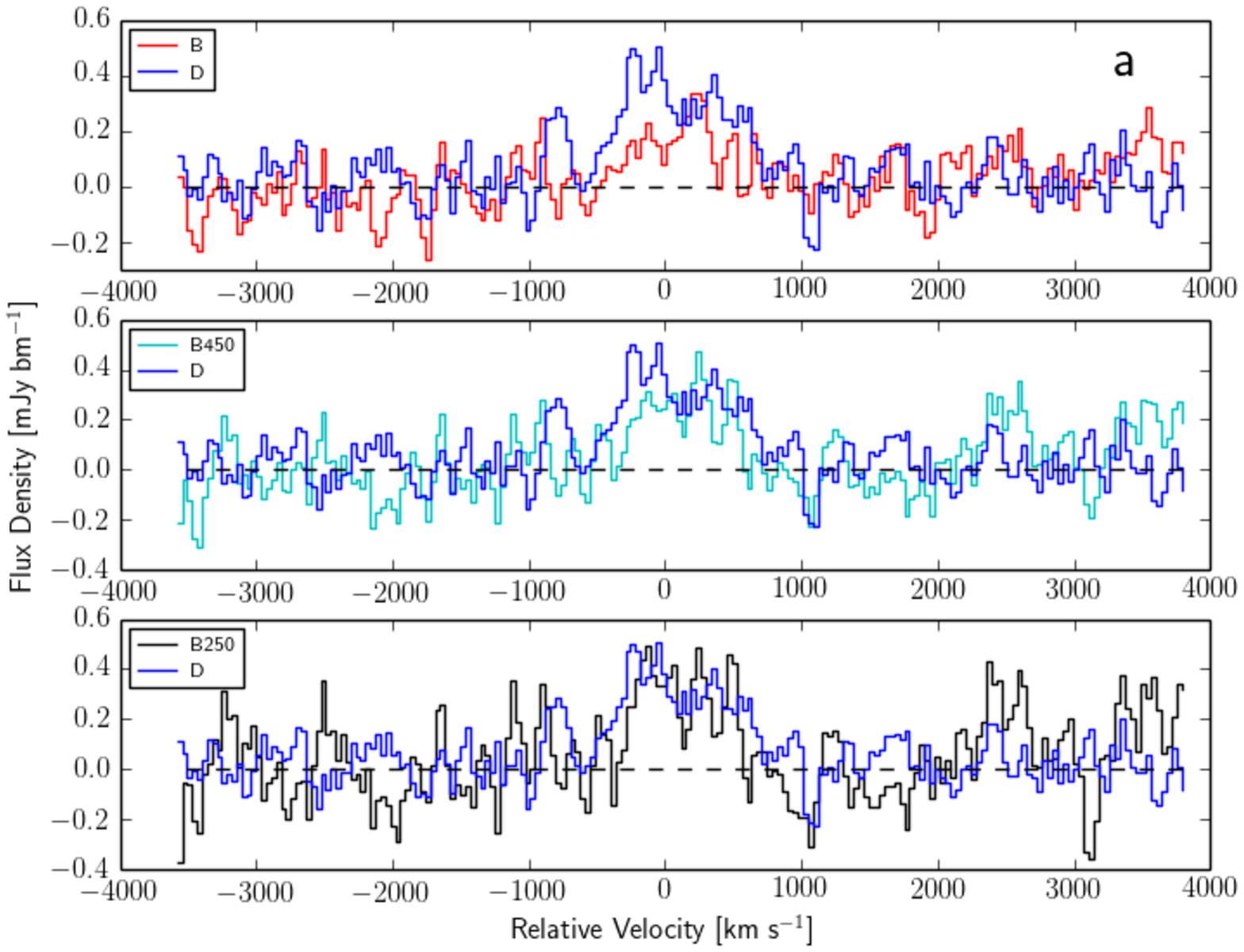}
\includegraphics[scale=0.5,clip=true]{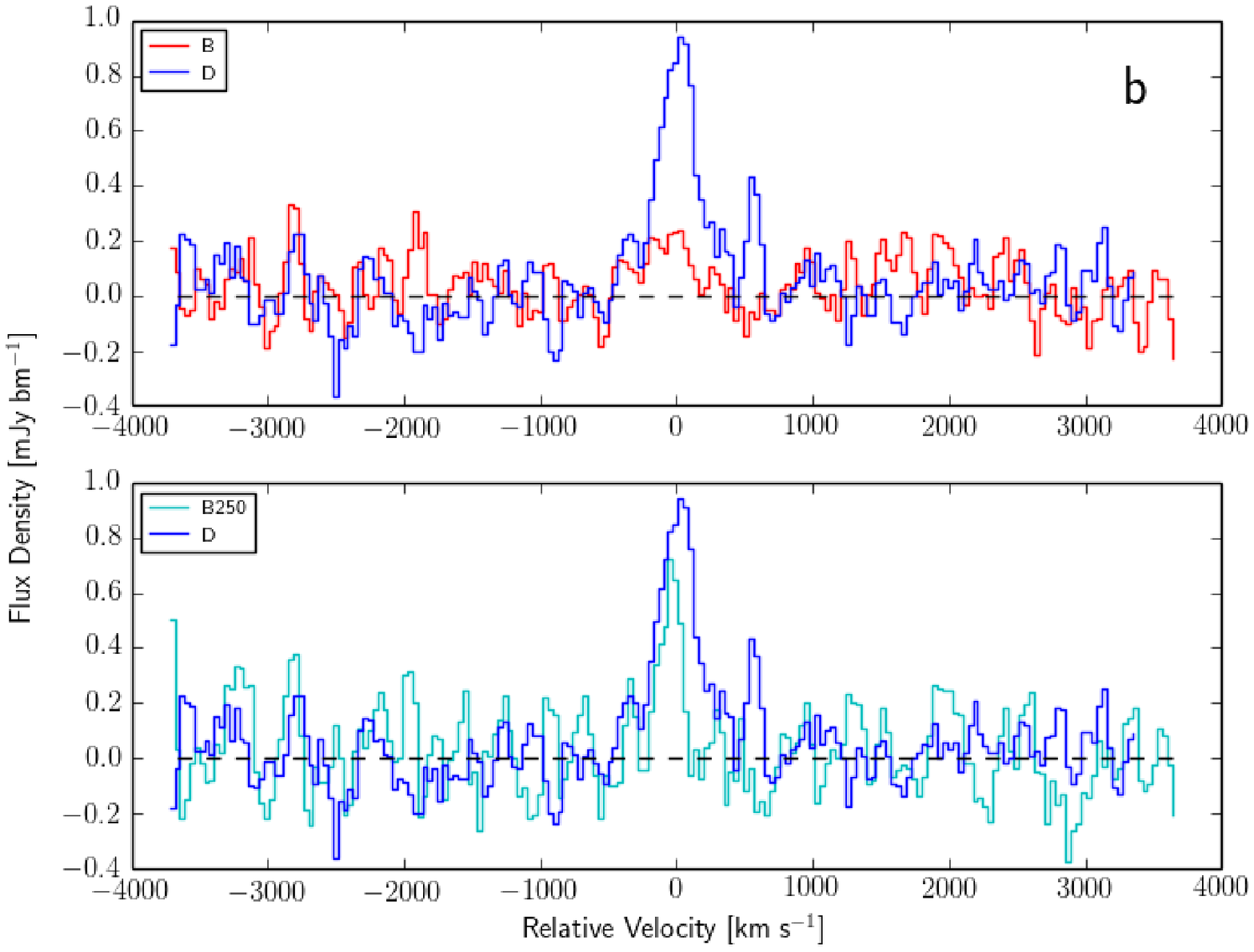}
\caption{Spectra taken at emission peaks of the SMG (a) and QSO (b) of BRI1202--0725. These are extracted from the following images: untapered B--configuration (B), untapered D--configuration (D), B--configuration tapered to 450k$\lambda$ (B450), and B--configuration tapered to 250k$\lambda$ (B250).}
\label{Nspec}
\end{figure}

All spectral line image cubes were created with natural weighting (after tapering), and
1$\,$GHz total bandwidth, approximately centered on the CO($J=2\rightarrow1$) line. Each channel was 5$\,$MHz wide, or $\sim35\,$km
s$^{-1}$. Fig. \ref{Nspec} compares spectra taken at the
emission peak of the untapered B--configuration, tapered
B--configurations (B750, B450, and B250), and D--configuration data
for the SMG and QSO of BRI1202--0725. The results of fitting
single 1-D Gaussians to spectra of each dataset taken at the B750 emission peaks are also
listed in Table \ref{north1}.

\begin{deluxetable}{lccccccc}
\tablecolumns{8}
\tablewidth{0pt}
\tabletypesize{\scriptsize}
\tablecaption{CO(2$\rightarrow$1) emission peak spectrum fits and image properties \label{north1}}
\tablehead{
  & & \colhead{Amplitude} & \colhead{Center} & \colhead{$\Delta$v$_{\tiny{FWHM}}$} & \colhead{Line Integral} & \colhead{Restoring Beam} & \colhead{$\sigma$}\\ 
  \colhead{Source} & \colhead{Image} & & & & & & \\
  & & \colhead{[mJy beam$^{-1}$]} & \colhead{[km s$^{-1}$]} & \colhead{[km s$^{-1}$]} & \colhead{[mJy beam$^{-1}$ km s$^{-1}$]} & \colhead{[$''\times ''$]} & \colhead{[mJy beam$^{-1}$]}}
\startdata
 BRI1202--0725 SMG & B & $0.23 \pm 0.02$ & $115 \pm 35$&$688 \pm 81$&$168 \pm 18$& 0.22$\times$0.16 &0.17\\
 BRI1202--0725 SMG & B750 & $0.28 \pm 0.03$ &$ 113 \pm 29 $&$ 685 \pm 69 $&$ 206 \pm 19$& 0.29$\times$0.21 &0.18\\
 BRI1202--0725 SMG & B450 & $0.31 \pm 0.03$&$99 \pm 35$&$724 \pm 83$&$236 \pm 25$& 0.38$\times$0.29 &0.20\\
 BRI1202--0725 SMG & B250 &$ 0.37 \pm 0.05 $&$ 55 \pm 46 $&$ 727 \pm 108 $&$ 286 \pm 40$& 0.58$\times$0.44 &0.26\\
 BRI1202--0725 SMG & D &$ 0.38 \pm 0.03 $&$ 9 \pm 37 $&$ 1052 \pm 88 $&$ 421 \pm 33$& 2.57$\times$1.73 & 0.18\\ \hline
 BRI1202--0725 QSO & B & $0.21 \pm 0.03$ & $-96 \pm 29$&$422 \pm 69$&$96 \pm 15$& 0.22$\times$0.16 &0.17\\
 BRI1202--0725 QSO & B750 & $0.37 \pm 0.04$ &$ -72 \pm 14 $&$ 262 \pm 33 $&$ 102 \pm 12$& 0.29$\times$0.21 &0.18\\
 BRI1202--0725 QSO & B450 & $0.50 \pm 0.04$&$-69 \pm 8$&$196 \pm 19$&$105 \pm 9$& 0.38$\times$0.29 &0.20\\
 BRI1202--0725 QSO & B250 &$ 0.70 \pm 0.06 $&$ -64 \pm 8 $&$ 174 \pm 18 $&$ 128 \pm 13$& 0.58$\times$0.44 &0.26\\
 BRI1202--0725 QSO & D &$ 0.87 \pm 0.06 $&$ -14 \pm 11 $&$ 365 \pm 27 $&$ 336 \pm 23$& 2.57$\times$1.73 & 0.18\\ \hline \hline
 BRI1335--0417 N & B & $0.18 \pm 0.03$ & $46 \pm 31$&$346 \pm 73$&$67 \pm 13$&0.23$\times$ 0.15& 0.13\\
 BRI1335--0417 N & B750 & $0.27 \pm 0.03$ &$77 \pm 25 $&$400  \pm  59$&$114 \pm 16$&0.30$\times$0.21&0.13\\
 BRI1335--0417 N & B450 & $0.34 \pm 0.04$&$94 \pm 21$&$412 \pm 49$&$150 \pm 17$& 0.36$\times$0.29&0.15\\
 BRI1335--0417 N & B250 &$0.49  \pm  0.04$&$ 103 \pm 15 $&$ 424 \pm  35$&$ 222 \pm 17$& 0.54$\times$0.46&0.19\\
 BRI1335--0417 N & D &$ 1.14 \pm 0.05 $&$ 40 \pm 6 $&$ 320 \pm 15 $&$ 387 \pm 17 $&2.54$\times$1.95&0.29\\ \hline
 BRI1335--0417 S & B & $0.30 \pm 0.02$ & $ 58\pm 16$&$486 \pm 37$&$153 \pm 11$&0.23$\times$ 0.15& 0.13\\
 BRI1335--0417 S & B750 & $0.33 \pm 0.02$ &$ 35 \pm 13 $&$ 537 \pm 31 $&$ 190 \pm 10$&0.30$\times$0.21&0.13\\
 BRI1335--0417 S & B450 & $ 0.38\pm 0.02$&$36 \pm 14$&$556 \pm 33$&$226 \pm13$& 0.36$\times$0.29&0.15\\
 BRI1335--0417 S & B250 &$ 0.51 \pm 0.03 $&$ 41 \pm 15 $&$ 503 \pm 36 $&$ 273 \pm 18$& 0.54$\times$0.46&0.19\\
 BRI1335--0417 S & D &$ 1.15 \pm 0.05$&$ 33\pm 7 $&$ 331 \pm 16 $&$ 404 \pm 18$&2.54$\times$1.95&0.29\\ \hline
\enddata
\end{deluxetable}

The first set of plots in Fig. \ref{Nspec} (a) shows that, for the
BRI1202--0725 SMG, the untapered B--configuration captures less that
half of the D--configuration line flux, and falls particularly short at negative
velocity.  The 250k$\lambda$ taper spectrum is comparable to that of
the D--configuration, within the noise, indicating a source size approximately equal to the beam size. From Table \ref{north1}, this
places the scale of the BRI1202--0725 SMG at around that of the
restoring beam of B250, namely $\sim0.5''$ ($\sim 3.3\,$kpc). 

The second set of plots (b) shows that the full B--configuration resolution
spectrum of the QSO recovers very little of the CO emission, and
essentially zero emission on the high positive velocity wing of the line.  The
B250 taper captures most of the negative velocity part of the emission, but
still misses most of the D--configuration emission above zero velocity,
implying an extent $> 0.5''$ ($> 3.3\,$kpc).

\begin{figure}[h]
\centering
\includegraphics[scale=0.5,clip=true]{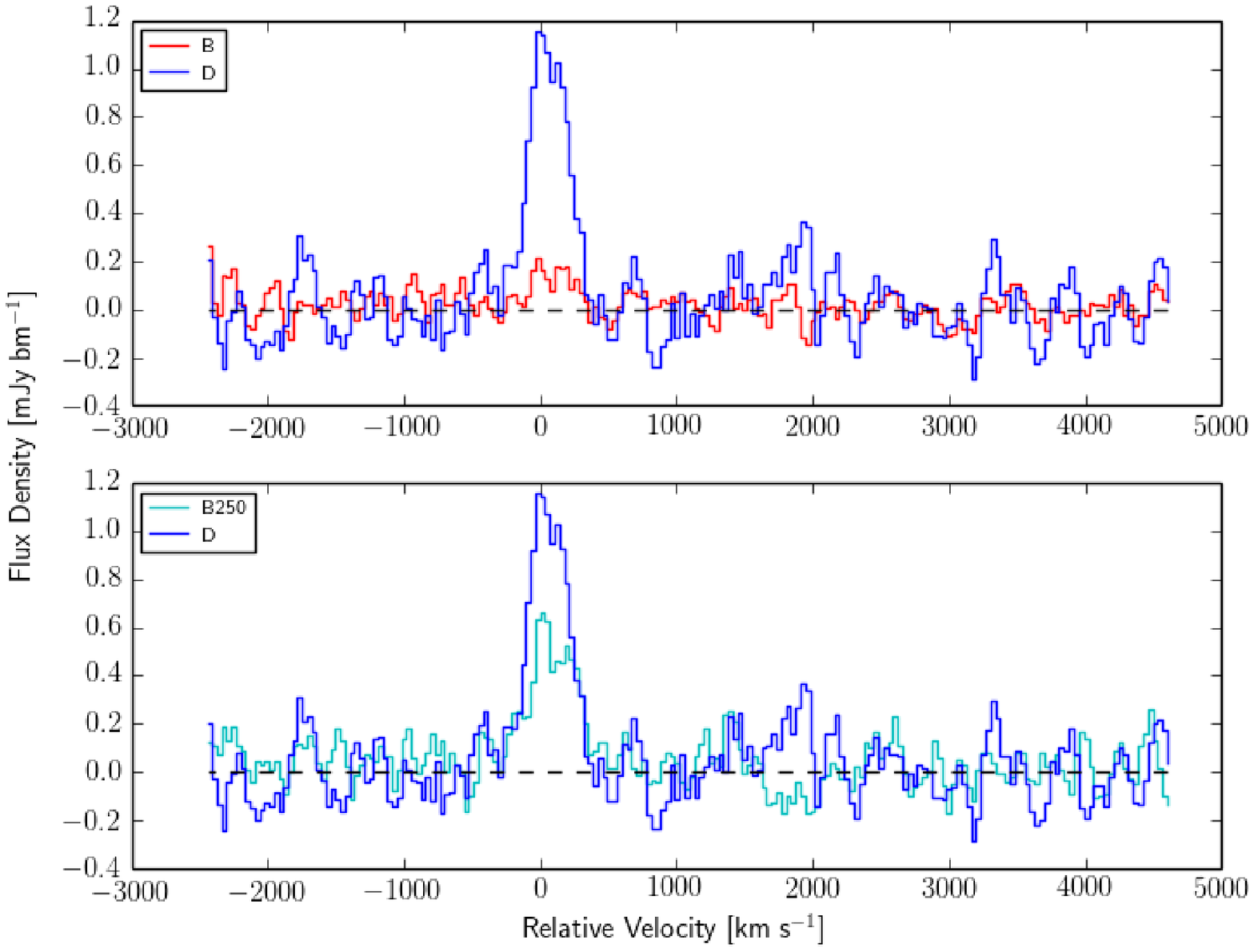}
\caption{Spectra taken at the emission peak of BRI1335--0417. These are extracted from the following images: untapered B--configuration (B), untapered D--configuration (D), and B--configuration tapered to 250k$\lambda$ (B250).}
\label{ooooo}
\end{figure}

BRI1335--0417 is an extended CO source, as was already
demonstrated in \citet{riec08}.  This is seen clearly in the
comparison of the full B--configuration resolution spectrum and the D--configuration (Fig. \ref{ooooo}).
Even for the B250 taper, we are only recovering about half of the
integrated D--configuration emission.  From this, we conclude that the extent of the emission 
is $>> 0.5''$ ($>> 3.4\,$kpc).

\subsection{Moment Zero Images}

Our spectral analysis has demonstrated clearly that the CO emission 
is spatially extended in all three sources.  We now turn
to the imaging analysis for further details, keeping in mind that
the signal to noise ratio for some of the sources is low, and hence
detailed imaging is difficult. We employ the B750 images, which retain
some brightness sensitivity, but also provide reasonable resolution ($\sim 0.25''$). 

Moment zero (velocity integrated CO flux) images were generated by applying the
CASA task \textit{immoments} to the tapered image cubes. The velocity ranges for
integration were determined by examining the width (Full Width at Zero Intensity; FWZI) of the CO lines in each dataset. This extent was
-357 to 606$\,$km s$^{-1}$ (40.43 to 40.56$\,$GHz) for the BRI1202--0725 SMG, -277 to 241$\,$km s$^{-1}$ (40.46 to 40.53$\,$GHz)
for the BRI1202--0725 QSO, and -281 to 392$\,$km s$^{-1}$ (42.595 to 42.690$\,$GHz) for BRI1335--0417.

\begin{figure}[h]
\centering
\includegraphics[scale=0.4,clip=true]{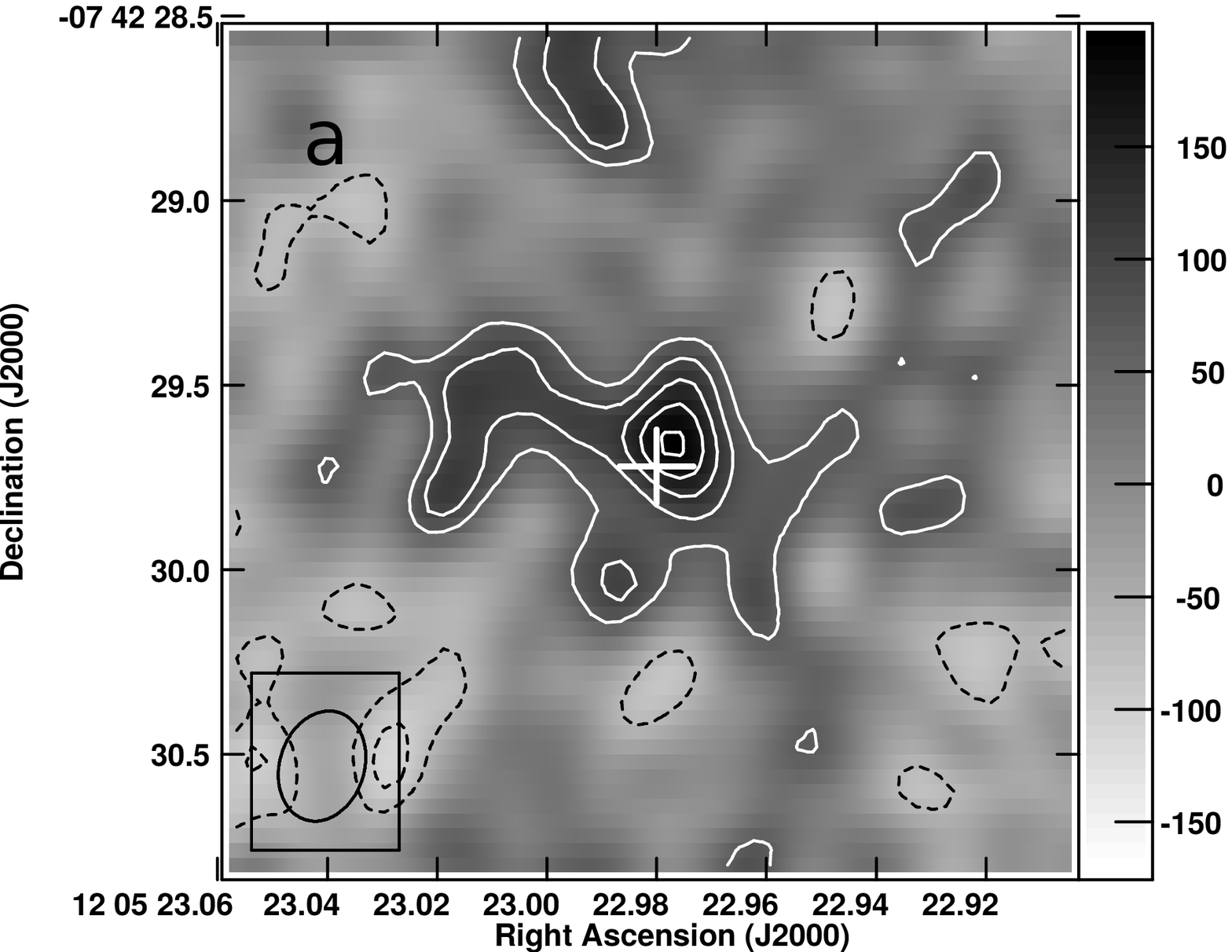}
\includegraphics[scale=0.4,clip=true]{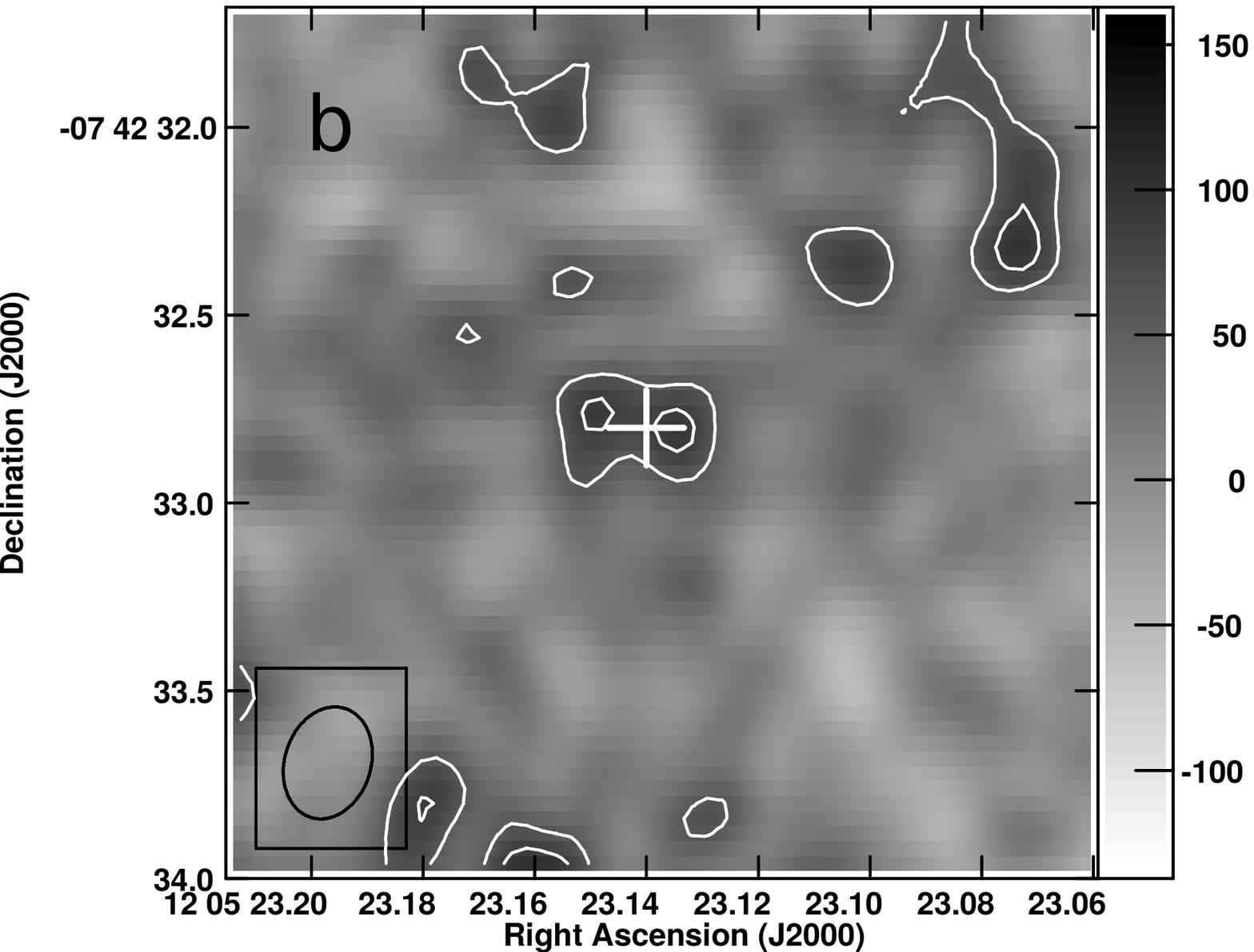}
\caption{Moment zero images of BRI1202--0725 made with the B--configuration data tapered to 750k$\lambda$ (B750), averaged over the width of the CO($J=2\rightarrow1$) line as seen in B750. Contours begin at $\pm2\sigma$ and are in steps of $1\sigma$. 44$\,$GHz continuum positions of this work shown by crosses (size of $\sim 0.2''$ corresponds to positional uncertainty). The restoring beam $(0.31''\times0.23'')$ at position angle $-17^{\circ}$ is shown as an ellipse in the lower left corner. a) Image of the SMG, $1\sigma= 32.3\,$mJy beam$^{-1}$ km s$^{-1}$. b) Image of the QSO, $1\sigma=28.8\,$mJy beam$^{-1}$ km s$^{-1}$. The greyscale is in units of mJy beam$^{-1}$ km s$^{-1}$.}
\label{mom}
\end{figure}

For the SMG in BRI1202--0725, the moment zero map in Fig. \ref{mom}a
show an east-west extension at the $3\sigma$ level, with a maximum
extent of $\sim 0.6''$. This extension is admittedly of marginal significance, but it is certainly consistent with the
direction of the velocity gradient seen in [CII] images
taken with ALMA (\citealt{wagg12,cari13,carn13}). 

For the QSO in BRI1202--0725, the B750 moment zero image (Fig. \ref{mom}b) shows a marginal (3$\sigma$) detection, with only a fraction of the integrated flux density seen in the D--configuration, implying that the B750 image resolves out
most of the emission. It should be noted that fits to each of the two components of the B750 data show that the emission is equally shared between them ($0.12\pm0.03\,$Jy km s$^{-1}$ for the west peak and $0.13\pm0.03\,$Jy km s $^{-1}$ for the east peak). For completeness, a single Gaussian fit to the QSO emission yields
$(3.1\pm0.3)\times(1.1\pm0.5)\,$kpc$^2$ at PA = 81$\pm3^{\circ}$, with a peak of $0.09\pm0.03\,$Jy beam$^{-1}$
km s$^{-1}$ and a total of $0.24\pm0.03\,$Jy km s $^{-1}$. This integrated flux density is
$75\%$ that of the D--configuration data ($0.32\pm0.01\,$Jy km s $^{-1}$; \citealt{wagg14}). 

%\begin{figure}[h]
%\centering
%\includegraphics[scale=0.4,clip=true]{f6.eps}
%\caption{Spectra of the BRI1202--0725 SMG taken over the beam of the D--configuration and over the combined core and extension of the B750 dataset.}
%\label{rot}
%\end{figure}

\begin{figure}[h]
\centering
\includegraphics[scale=0.28,clip=true]{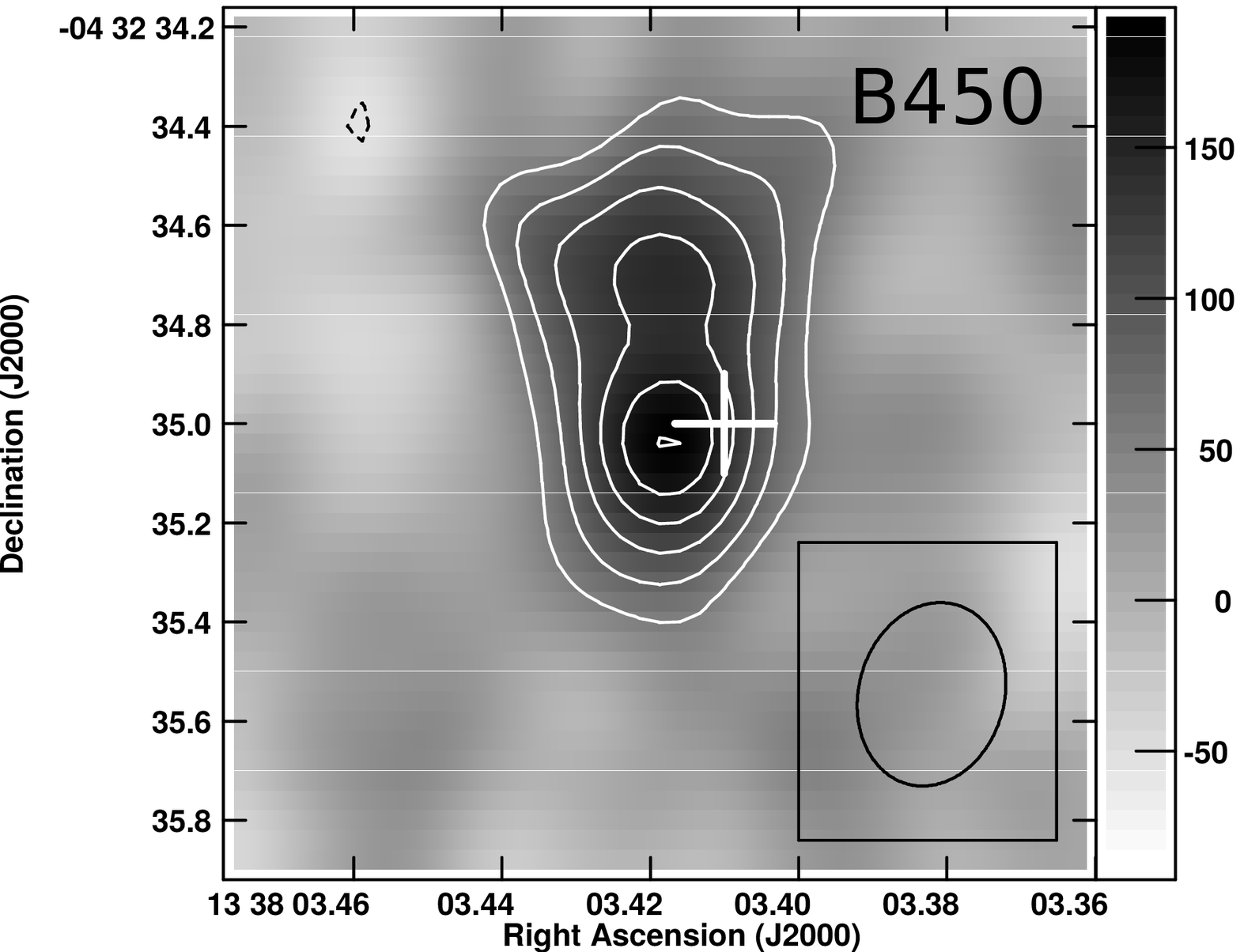}
\includegraphics[scale=0.28,clip=true]{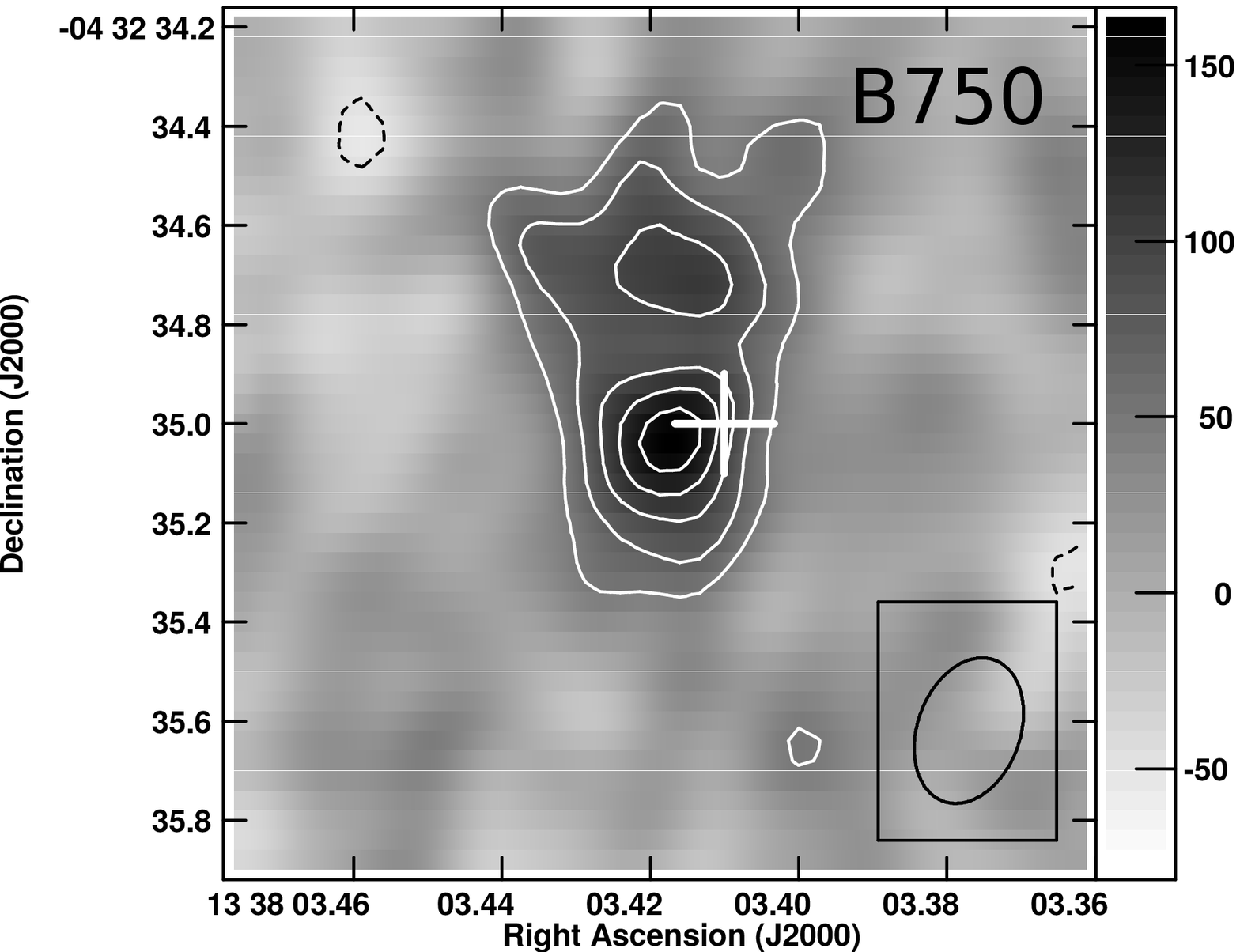}
\includegraphics[scale=0.28,clip=true]{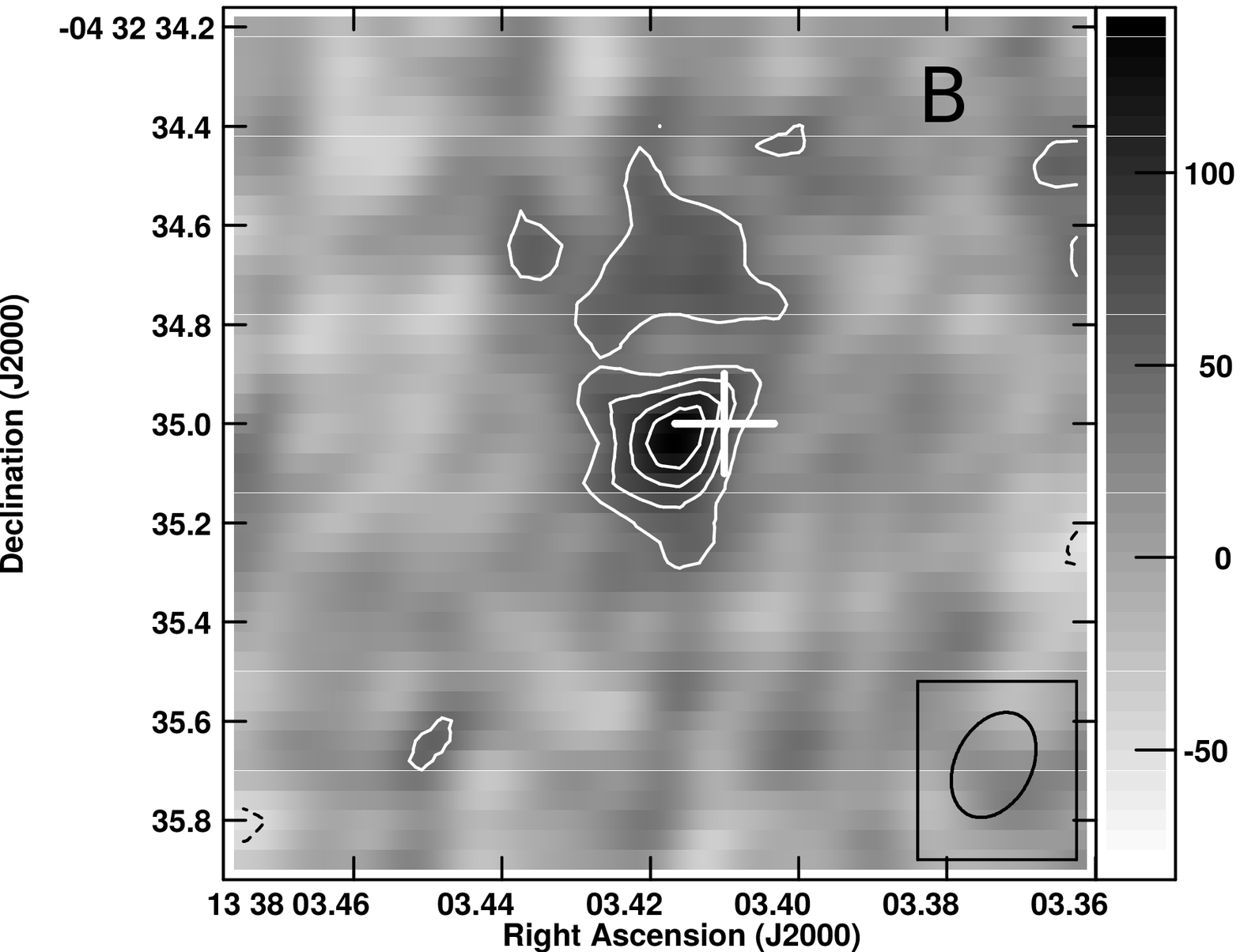}
\caption{Moment zero images of BRI1335--0417 made with tapered and untapered B--configuration data, integrated over the width of the CO($J=2\rightarrow1$) line. Location of continuum emission from this work shown by cross (size of $\sim0.2''$ corresponds to positional uncertainty based on our beam size). Contours begin at $\pm2\sigma$ and are in steps of $1\sigma$. Restoring beams shown in lower right corner. The greyscale is in units of mJy beam$^{-1}$ km s$^{-1}$.
Left: B--configuration tapered to 450k$\lambda$, $1\sigma=27\,$mJy beam$^{-1}$ km s$^{-1}$. Restoring beam is $(0.38''\times0.29'')$ at position angle $-15^{\circ}$. 
Middle: B--configuration tapered to 750k$\lambda$, $1\sigma=24\,$mJy beam$^{-1}$ km s$^{-1}$. Restoring beam is $(0.30''\times0.21'')$ at position angle $-20^{\circ}$.
Right: Untapered B--configuration, $1\sigma=23\,$mJy beam$^{-1}$ km s$^{-1}$. Restoring beam is $(0.23''\times0.15'')$ at position angle $-27^{\circ}$.}
\label{1335mom0}
\end{figure}

Moment zero images of BRI1335--0417 (Fig. \ref{1335mom0}) show
clearly the extended emission to the north and the ever present dominant southern source.  We have generated 
moment zero images at three different resolutions, showing the emergence
of diffuse structure to the north. Fitting Gaussians to the southern and northern sources separately in the B750 image yields sizes of $(2.2\pm0.1)\times(1.8\pm0.2)\,$kpc$^2$ at PA = $(26\pm20)^{\circ}$ and $(2.1\pm0.1)\times(0.8\pm0.2)\,$kpc$^2$ at PA = $(71\pm5)^{\circ}$, respectively. A fit to the brighter southern source gives a larger integrated flux density of ($0.36\pm0.02)\,$Jy km s$^{-1}$ versus ($0.21\pm0.02)\,$Jy km s$^{-1}$ for the northern source. It also features a higher fit peak flux ($0.15\pm0.02)\,$Jy beam$^{-1}$ km s$^{-1}$ versus ($0.11\pm0.02)\,$Jy beam$^{-1}$ km s$^{-1}$ for the northern source. A fit to the D--configuration data returns an integrated flux density of $(0.62\pm0.03)\,$Jy km s$^{-1}$, which is comparable to the two above B750 fits combined ($0.57\pm0.04\,$Jy km s$^{-1}$). This suggests that the majority of the CO emission emanates from these two compact sources, not from diffuse gas.

\section{Analysis}

\subsection{BRI1202--0725: Ly$\alpha$-2 CO Emission}

\begin{figure}[h]
\centering
\includegraphics[scale=0.5,clip=true]{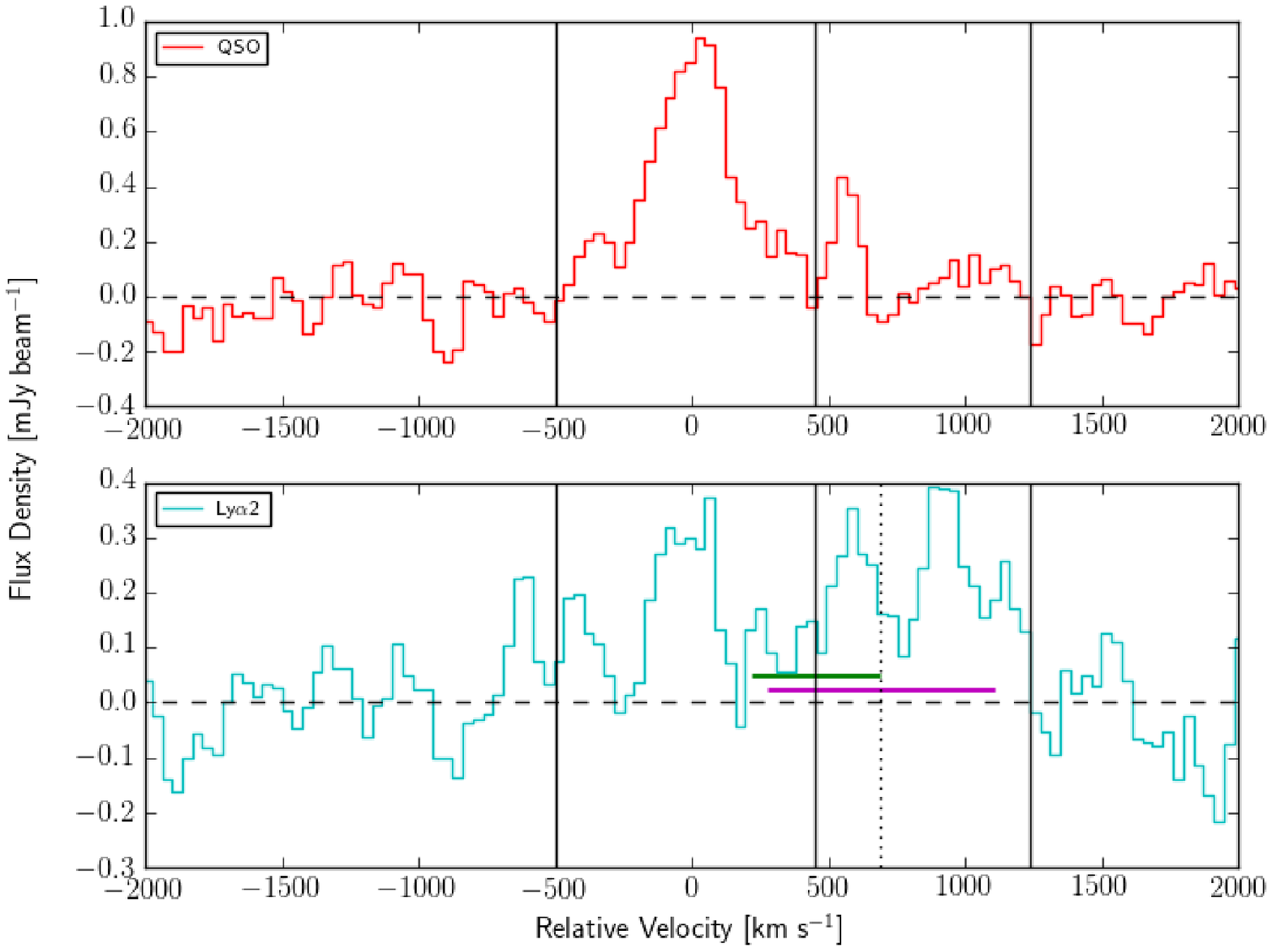}
\caption{Comparison of D--configuration peak spectra of the BRI1202--0725 QSO and Ly$\alpha$-2. Solid vertical lines divide channels of Ly$\alpha$-2 emission from those of QSO emission. Green and magenta lines in lower panel shows centroid and FWTM of [CII] from \citet{cari13} and [NII] emission from \citet{deca14}, respectively. Note that the width of the [CII] line is cutoff at the observed band edge, as shown by dotted vertical line.}
\label{1202twospec}
\end{figure}

\begin{figure}[h]
\centering
\includegraphics[scale=0.4,clip=true]{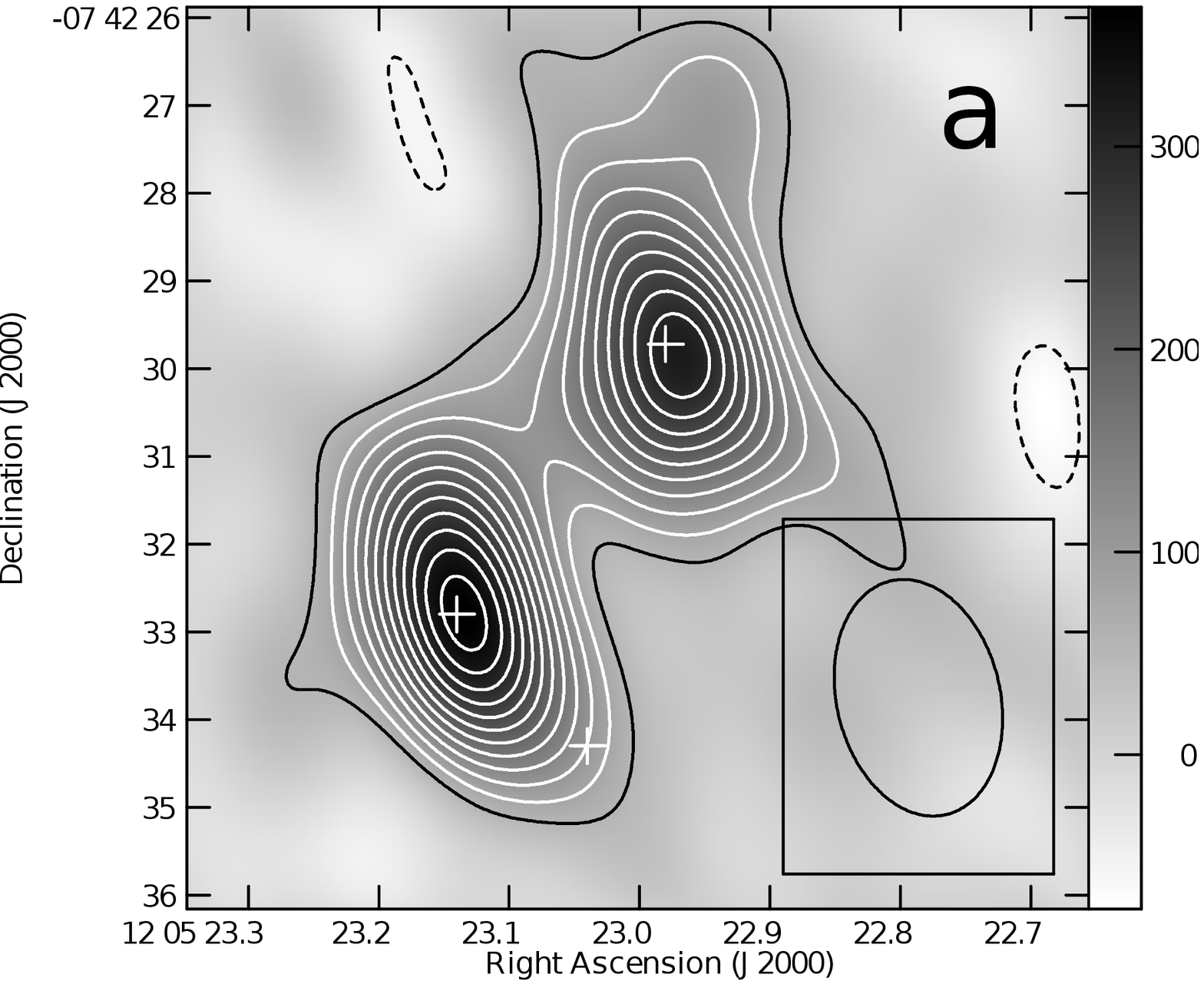}
\includegraphics[scale=0.39,clip=true]{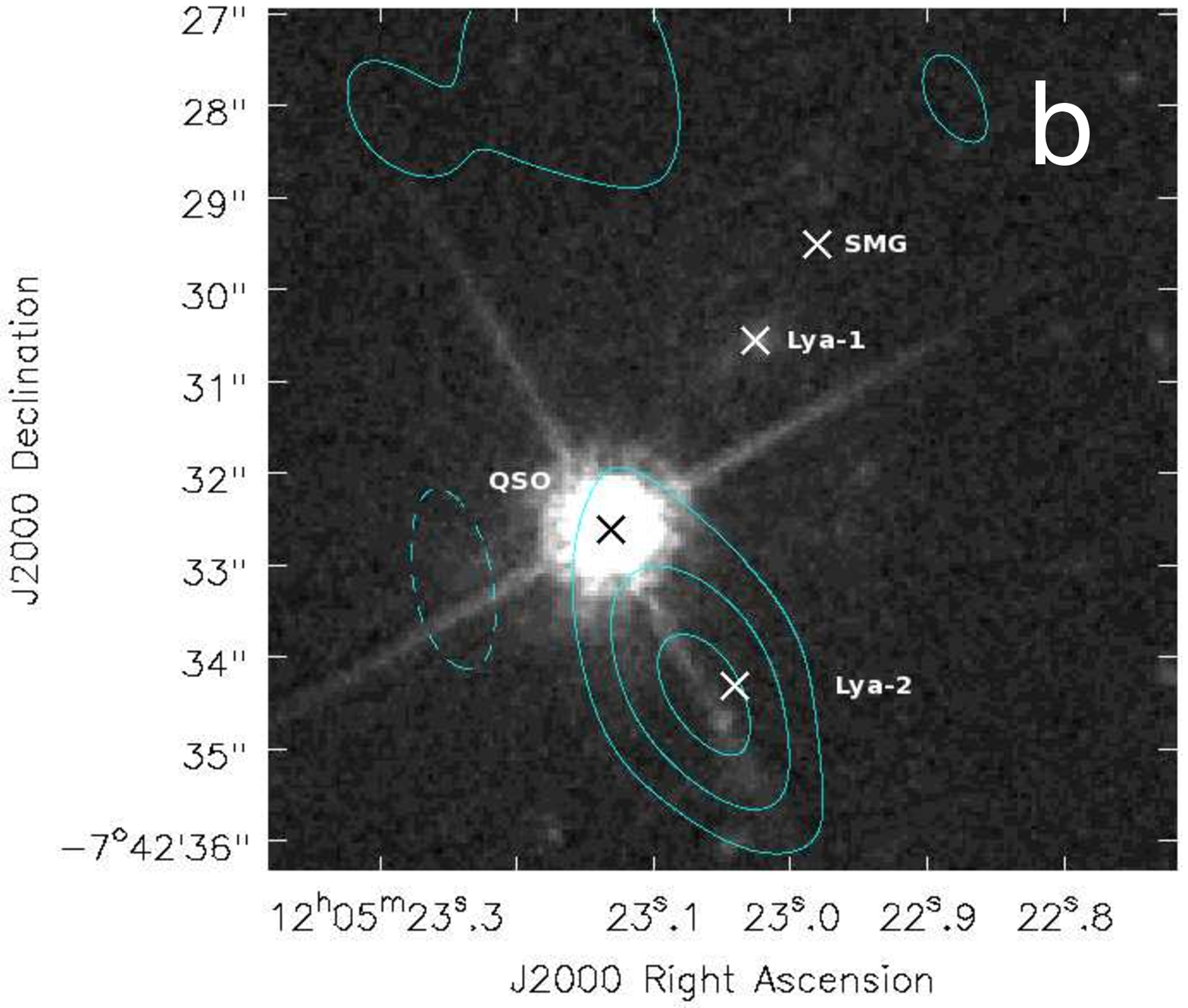}
\caption{Moment zero images of the BRI1202--0725 field from D--configuration data, integrated over the QSO and Ly$\alpha$-2 linewidths.  
a) The field integrated over the QSO linewidth (-500 to 450$\,$km s$^{-1}$; 40.4209 to 40.5491$\,$GHz). The crosses represent (from north to south) the 44$\,$GHz continuum positions of the SMG and QSO from this work and the \citet{cari13} submillimeter positions of Ly$\alpha$-2 (cross sizes are arbitrary). Contours begin at $\pm2\sigma$ and are in steps of $1\sigma= 27\,$mJy beam $^{-1}$ km s$^{-1}$. Restoring beams shown as an ellipse in the lower right corner. The greyscale is in units of mJy beam$^{-1}$ km s$^{-1}$. The restoring beam is $(2.74''\times1.86'')$ at position angle $14^{\circ}$. 
b) The field integrated over the Ly$\alpha$-2 linewidth (450 to 1240$\,$km s $^{-1}$; 40.5491 to 40.6563$\,$GHz). Contours begin at $\pm2\sigma$ and are in steps of $2\sigma= 30\,$mJy beam $^{-1}$ km s$^{-1}$. Positions of 340$\,$GHz detections of \citet{cari13} are marked. The color shows the \textit{HST}/ACS F775W filter image of \citet{deca14}}
\label{food}
\end{figure}

We next investigate CO emission from Ly$\alpha$-2 to the
southwest of the QSO using the D--configuration observations (Fig. \ref{1202twospec} and \ref{food}). Velocities are relative to the QSO redshift.
Ly$\alpha$-2 is separated by $2.5''$ from the QSO, corresponding to less than the 20\%
point of the synthesized beam. Note that no CO emission was detected from Ly$\alpha$-1 in any dataset.

Fig. \ref{1202twospec} shows the CO($J=2\rightarrow1$) spectrum of the QSO, again,
as well as one taken at the position of Ly$\alpha$-2. Two velocity ranges are
indicated: one corresponding to the main QSO line (-500 to 450$\,$km s$^{-1}$),
and a second at higher positive velocity (450 to 1240$\,$km s$^{-1}$).
Note that the main QSO line has no emission over most of this high positive
velocity range, hence sidelobes from the QSO emission are not an issue in terms of contaminating
the Ly$\alpha$-2 spectrum.  There appears to be emission over this high positive
velocity range at the position of Ly$\alpha$-2. The low end of this 
velocity range also corresponds to the [CII] emission seen from Ly$\alpha$-2 \citep{wagg12}. 
Unfortunately, the ALMA [CII] observations covered only the lower
velocities. The full width at tenth of maximum (FWTM) of the [NII] observation of \citet{deca14} (magenta line) is comparable to our upper velocity bin.

Fig. \ref{food} shows an image of the integrated emission in the high
positive velocity range.  A clear source appears at the
position of Ly$\alpha$-2, with a 6$\sigma$ significance.  The
integrated line flux density over this velocity range is
$0.19\pm0.03\,$Jy km $s^{-1}$, which is significantly larger than the
limit of $<0.06\,$Jy km $s^{-1}$ from \citet{wagg14}.  Adopting
bounding conversion factors of
$\alpha_{CO}=0.8$M$_{\odot}$K$^{-1}$ km$^{-1}$ s pc$^{-2}$ (for
starbursting galaxies) and $\alpha_{CO}=4$M$_{\odot}$K$^{-1}$
km$^{-1}$ s pc$^{-2}$ (for the Milky Way) yields a gas mass range of
($3.2-16)\times10^{10}\,$M$_{\odot}$.

Briefly, we consider why Ly$\alpha$-2 CO(J=$2\rightarrow1$)
emission was not reported in the original paper of \citet{wagg14}.
First, Wagg et al. were focused principally on the BRI1202--0725 SMG and AGN host
galaxies.  And second, the velocity selection of Ly$\alpha$-2 was based on
the truncated [CII] line of \citet{cari13}.  Recent [NII] observations
by \citet{deca14} show that the full profile of the object is
much broader.  We have increased the signal-to-noise by opening up the search
space in velocity to accommodate a broader line. 

We have looked for the emission from Ly$\alpha$-2 CO in the B--configuration data
over the broader velocity range. We do not detect emission at
full resolution, of B750, implying a $3\sigma$ level of $0.1\,$Jy beam$^{-1}$ km s$^{-1}$. These results give a 
rough lower limit to the source size of $>0.3"$.  Conversely, fitting to the D array
image of the velocity integrated CO emission leads to an upper limit to the
size of $<3''$. We may then state a conservative CO extent of $2-20\,$kpc.

\subsection{Star Formation Surface Densities}
 
We have SFRs for each object \citep{wagg14}, but require star
formation areas to calculate surface densities. Current measurements
are only limits or marginal measurements. \citet{carn13} present
unresolved ALMA 340$\,$GHz size estimates of $<22\,$kpc$^2$ for the
BRI1202--0725 SMG and QSO. Our measured 44$\,$GHz size for the
BRI1202--0725 SMG was $0.9\pm0.5\,$kpc$^2$, which is an order of
magnitude smaller. Another constraint is the 1.4$\,$GHz size of the
BRI1202--0725 SMG from \citet{momj05},
$(0.3\pm0.2)\,$kpc$^2$. Similarly, we found that BRI1335--0417 showed
a 44$\,$GHz size of $1.8\pm1.2\,$kpc$^2$, while \citet{momj07} found a
1.4$\,$GHz size of $(0.4\pm0.2)\,$kpc$^2$. These 1.4GHz
synchrotron-emitting areas are smaller than, but within
$\sim1.2\sigma$ of our $44\,$GHz regions. While our 44$\,$GHz size
estimates feature large uncertainties, they also show integrated flux
densities that are greater than their peak surface brightnesses
($41\pm6\,\mu$Jy and $23\pm6\,\mu$Jy beam$^{-1}$ for BRI1202--0725
SMG, $24\pm5\,\mu$Jy and $10\pm3\,\mu$Jy beam$^{-1}$ for
BRI1335--0417). The greater integrated flux density vs. peak surface
brightness suggests that the sources are indeed resolved.  However, in
both cases, these integrated flux densities from the B array imaging
are still (marginally) smaller than those seen in the D array data:
$51 \pm 6$ $\mu$Jy and $40 \pm 7$ $\mu$Jy, respectively, suggesting
additional diffuse emission. Given the low signal-to-noise of these
observations, we adopt for the sake of calculation the 44GHz sizes as
as a guide, keeping in mind that these are likely lower limits, with
upper limits set by ALMA dust continuum imaging. 

The BRI1202--0725 SMG shows a density of
$\Sigma_{SFR}=(2\pm1)\times10^3\,$M$_{\odot}$ yr$^{-1}$ kpc$^{-2}$,
while BRI1335--0417 shows
$\Sigma_{SFR}=(3\pm2)\times10^3\,$M$_{\odot}$ yr$^{-1}$
kpc$^{-2}$. The continuum image of the BRI1202--0725 QSO has too low a
signal to noise ratio to determine a size. The SFR surface density of
Ly$\alpha$-2 may be estimated using a SFR=$170\,$M$_{\odot}$ yr$^{-1}$
\citep{cari13} and a 44$\,$GHz size limit of $>1.4''\sim9\,$kpc, giving
$<121\,$M$_{\odot}$ year$^{-1}$ kpc$^{-2}$.

The fraction of the luminosity at which gravitational collapse is balanced by radiation pressure, or Eddington fraction, of these sources can be estimated using these surface densities. \citet{thom05} develop the limit of SFR surface density $\Sigma_{SFR}>1000\,$M$_{\odot}$ year$^{-1}$ kpc$^{-2}$ for super-Eddington luminosity, assuming a star formation efficiency and opacity. By this criterion, both the BRI1202--0725 SMG and BRI1335--0417 are super-Eddington. An alternate estimate may be found using a luminosity-to-mass limit of $>$500$\,$L$_{\odot}$/M$_{\odot}$ \citep{scov12}. Approximate values of $10^{13}\,$L$_{\odot}$ and $10^{11}\,$M$_{\odot}$ give Eddington fractions of $\sim20\%$. See Appendix A for an additional Eddington criterion which returns $<1\%$ for both sources.

While the scatter is significant between these methods, the first suggests that both the SMG and BRI1335--0417 are radiating near their Eddington limit. It should be noted that the Eddington fraction is dependent on assumed geometry (disk/sphere), L'$_{CO}$ to M$_{H_2}$ conversion factor, opacity, and total object mass, among other variables.

\subsection{Gas Surface Densities}

For the sources that we have both 44$\,$GHz continuum and CO($J=2\rightarrow1$) size estimates (BRI1202--0725 SMG and BRI1335--0417), it is evident that the CO extent is greater. This continuum, which is $\sim250\,$GHz (rest frame), is a tracer of cold dust, and thus indirect star formation. The more compact size of the continuum emitting region compared to CO emission, which traces the molecular gas, suggests a central, excited region surrounded by a gaseous reservoir.

We estimate the molecular gas masses using three conversion factors from L$'_{CO}$ to M$_{H_2}$. However, all three factors use the luminosity of the $J=1-0$ transition of CO, not the observed $J=2-1$ transition. The ratios of these luminosities are given by \citet{cariw13} as r$_{21}$=L'$_{CO(2-1)}$/L'$_{CO(1-0)}$=0.85, 0.99, 0.97, and 0.5 for SMGs, QSOs, color selected galaxies (CSGs), and the Milky Way, respectively. The application of the SMG ratio to the BRI1202--0725 SMG and the QSO ratio to the BRI1202--0725 QSO are obvious. Less so is the correct correction for BRI1335--0417, which shows evidence of an AGN but no compact radio emission. We will use the SMG correction factor, due to its extended starburst region. To account for the unknown nature of BRI1202--0725 Ly$\alpha$-2, we will use the range of factors from Milky Way-type objects to CSGs (0.5-0.97).

The first gas mass estimate uses the standard conversion for low $z$ ULIRGs, $\alpha_{CO}=0.8\,$M$_{\odot}$K$^{-1}$ km$^{-1}$ s pc$^{-2}$. We also use the flexible factor of \citet{nara12}:
\begin{equation}
\alpha_{CO}=\frac{min[6.3,10.7\times W_{CO}^{-0.32}]}{Z'^{0.65}}
\label{eqnDN}
\end{equation}
where $Z'=Z/Z_{\odot}$ is the normalized metallicity and $W_{CO}$ is the line intensity, which we will define as $L'_{CO}/A_{CO}$, where $A_{CO}$ is the area of our CO($J=2-1$) emission and $L'_{CO}$ is our $L'_{CO(2-1)}/r_{21}$. To explore the metallicity dependence of the gas mass, we used solar metallicity ($Z'=1$) and the extremely low metallicity of I Zw 18 ($Z'=0.02$; \citealt{lequ79}). Note that the latter value is unphysically low for areas showing this level of stellar development, and it is only included as an illustrative boundary.

\begin{deluxetable}{lcccc}
\tablecolumns{5}
\tablewidth{0pt}
\tabletypesize{\scriptsize}
\tablecaption{Source Properties \label{allthat}}
\tablehead{
  \colhead{} & \colhead{BRI1202--0725 SMG} & \colhead{BRI1202--0725 QSO} & \colhead{BRI1335--0417} & \colhead{BRI1202--0725 Ly$\alpha$-2}}
\startdata
L'$_{CO(2-1)}$ [$10^{11}\,$K km s$^{-1}$ pc$^2$] & $1.1\pm0.2$ & $0.51\pm0.06$ & $1.09\pm0.08$ & $0.40\pm0.06$ \\
$\alpha_{CO}(Z'=1)$ [M$_{\odot}$ K$^{-1}$ km$^{-1}$ s pc$^{-2}$] & $0.41\pm0.05$ & $0.46\pm0.07$ & $0.40\pm0.02$ & $0.41-2.2$ \\
$\alpha_{CO}(Z'=0.02)$ [M$_{\odot}$ K$^{-1}$ km$^{-1}$ s pc$^{-2}$] & $5.2\pm0.6$ & $5.8\pm0.9$ & $5.1\pm0.2$ & $5.3-28$ \\
M$_{\alpha=0.8}$ [$10^{11}\,$M$_{\odot}$] & $1.0\pm0.2$ & $0.41\pm0.05$ & $1.03\pm0.07$ & $0.3-0.7$ \\
M$_{Z'=1}$ [$10^{10}\,$M$_{\odot}$] & $5\pm1$ & $2.3\pm0.5$ & $5.1\pm0.4$ & $1.5-3.3$ \\
M$_{Z'=0.02}$ [$10^{11}\,$M$_{\odot}$] & $7\pm1$ & $3.0\pm0.6$ & $6.5\pm0.5$ & $1.9-4.2$ \\
A$_{CO(2-1)}$ [kpc$^2$] & $5\pm2$ & $3\pm1$ & $4.4\pm0.5$ & $3-300$ \\
$\Sigma_{H_2,\alpha=0.8}$ [$10^4\,$M$_{\odot}$ pc$^{-2}$] & $2.1\pm0.8$ & $1.5\pm0.7$ & $2.3\pm0.3$ & $0.009-2$ \\
$\Sigma_{H_2,Z'=1}$ [$10^4\,$M$_{\odot}$ pc$^{-2}$] & $1.1\pm0.4$ & $0.9\pm0.4$ & $1.2\pm0.2$ & $0.005-1$ \\
$\Sigma_{H_2,Z'=0.02}$ [$10^5\,$M$_{\odot}$ pc$^{-2}$] & $1.4\pm0.6$ & $1.1\pm0.6$ & $1.5\pm0.2$ & $0.006-1$ \\
SFR [$10^3\,$M$_{\odot}$ year$^{-1}$]\tablenotemark{a} & $4\pm2$ & $4\pm2$ & $5\pm1$ & 0.17 \\
A$_{44\,\mathrm{GHz}}$ [kpc$^2$] & $0.9\pm0.5$ & - & $1.8\pm1.1$ & $>1.4$ \\
$\Sigma_{SFR}$ [$10^3\,$M$_{\odot}$ year$^{-1}$ kpc$^{-2}$] & $4\pm3$ & - & $3\pm2$ & $<120$ \\
\hline
\enddata
\tablenotetext{a}{\citet{wagg14}}
\end{deluxetable}

Our molecular gas masses and resulting surface densities are listed in Table \ref{allthat}. Using our CO emission extents (Section 4.3) and the molecular gas masses, we find gas surface densities $\Sigma_{H_2}\sim10^{4-5}\,$M$_{\odot}$ pc$^{-2}$ for the three HyLIRGs. These are larger than the average gas surface density reported by \citet{grev05}, who observed 11 SMGs in $z=1.0-3.4$ and found $\Sigma_{H_2}=(2.4\pm1.3)\times 10^3\,$M$_{\odot}$ pc$^{-2}$. Our values are more comparable to the estimates of \citet{dadd09}, who find $\Sigma_{H_2}=(1.6,1.0)\times 10^4\,$M$_{\odot}$ pc$^{-2}$ for GN20 and GN20.2a. 

\begin{figure}[h]
\centering
\includegraphics[scale=0.5,clip=true]{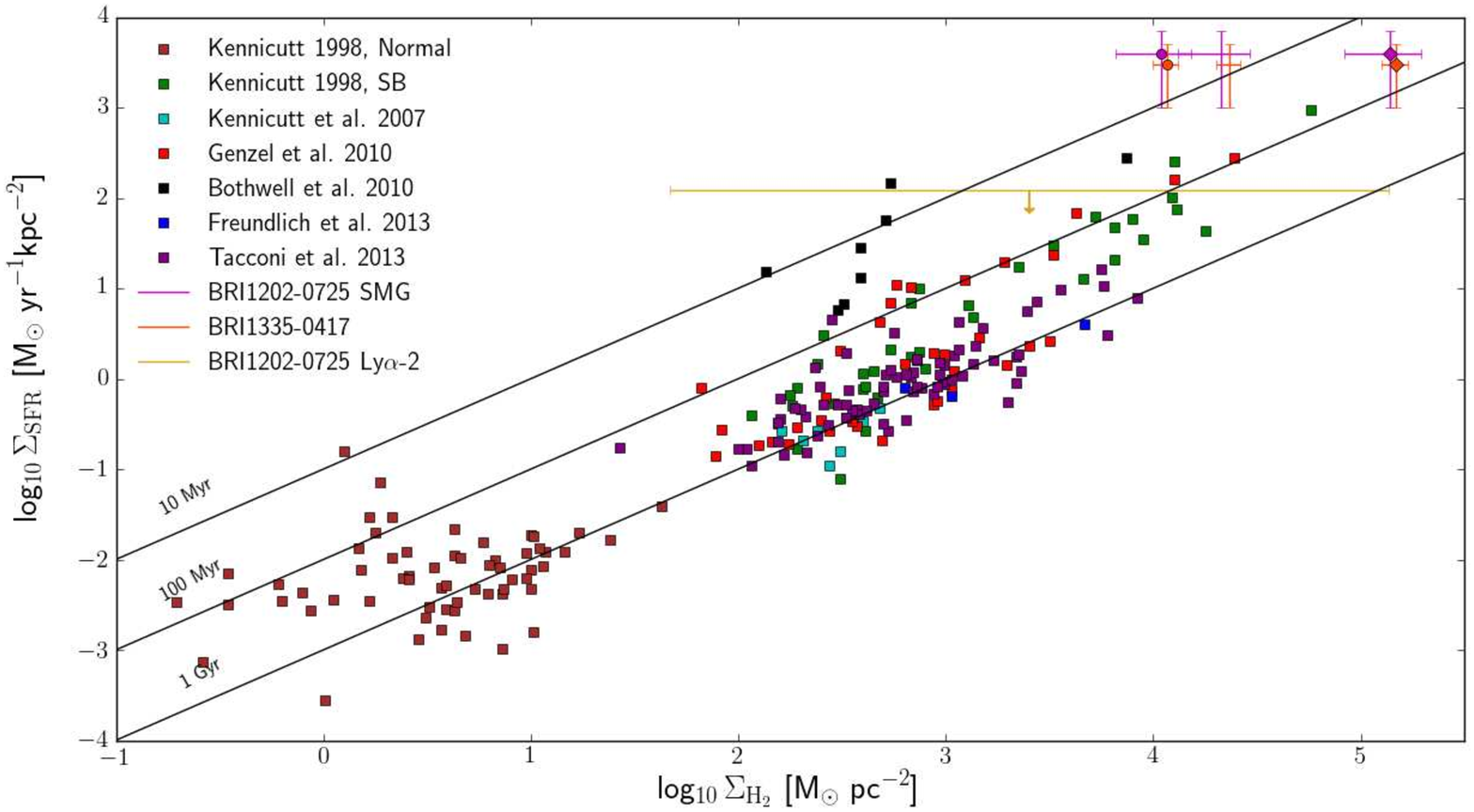}
\caption{Kennicutt-Schmidt diagram including our sources and those from literature. BRI1202--0725 QSO is not shown due to its unresolved 44$\,$GHz extent and resultant lack of $\Sigma_{SFR}$. Literature sources include normal galaxies (\citealt{kenn98,kenn07}, circumnuclear starburst areas (\citealt{kenn98}), $z\sim2$ SMGs (\citealt{genz10,both10}), $z\sim1$ SFGs (\citealt{freu13}), and $z\sim1-3$ SFGs of \citealt{tacc13}). Positions of our sources are shown using $\alpha_{CO}=0.8\,$M$_{\odot}$K$^{-1}$ km$^{-1}$ s pc$^{-2}$ (no marker), the adaptive factor of \citet{nara12} using $Z'=1$ (circles) and $Z'=0.02$ (diamonds).}
\label{KSPLOT}
\end{figure}

The placement of these objects on a Kennicutt-Schmidt plot (Fig. \ref{KSPLOT}) shows that they are separate from the main sequence. If an unrealistic metallicity is assumed, they may have comparable gas depletion timescales to local starbursts or $z\sim2$ SMGs. A limit on source metallicity can be be found by assuming $\alpha=0.8$  and inverting equation \ref{eqnDN} to solve for Z', yielding Z'$_{BRI1202-0725 SMG}=0.38\pm0.06$ and Z'$_{BRI1335-0417}=0.34\pm0.02$.

Finally, we may use our CO size limits and gas mass estimate for Ly$\alpha$-2 (Section 5.1) to find $\Sigma_{H_2}=10^{1.7-5.0}\,$M$_{\odot}$ pc$^{-2}$. This range includes our uncertainties in the CO size, the $\alpha_{CO}$ conversion factor, and the $r_{21}$ ratio. Due to its unconstrained size, it may match well with either starbursts or normal galaxies in the Kennicutt-Schmidt diagram (Fig. \ref{KSPLOT}).

\section{Discussion}

We have imaged the archetypal HyLIRGs BRI1202--0725 and BRI1335--0417
with the VLA B--configuration in the rest-frame 250$\,$GHz continuum
and CO($J=2\rightarrow1$) at high resolution. These observations allow us to determine
sizes for the 44GHz continuum and the CO emission
on scales down to $\sim 1$kpc.

The 44$\,$GHz continuum emission in all three sources appears extended on a scale
of 1 to 2$\,$kpc, although only marginally so for the QSO host in
BRI1202--0725.  Based on radio through FIR SED fitting,
the observed 44$\,$GHz continuum emission is thermal emission from
cold dust \citep{wagg14}. For the BRI1202--0725 SMG, the 44$\,$GHz size roughly agrees with the nonthermal 1.4$\,$GHz size from the VLBI observations of \citet{momj05}, while the 44$\,$GHz size of BRI1335--0417 is about twice as large as the VLBI 1.4$\,$GHz extent of \citet{momj07}. Assuming these 44$\,$GHz continuum extents correspond to that of the starbursting regions, we derive SFR surface densities of $\Sigma_{SFR}=(4\pm3)\times10^3\,$M$_{\odot}$ yr$^{-1}$ kpc$^{-2}$ for the BRI1202--0725 SMG and $\Sigma_{SFR}=(3\pm2)\,$M$_{\odot}$ yr$^{-1}$ kpc$^{-2}$ for BRI1335--0417. While estimates of the Eddington fractions of these sources vary, evidence from one test suggests that both of these objects are possibly radiating above their stable limit.

Using both the standard ULIRG conversion factor for CO luminosity to H$_2$ mass and the more flexible factor of \citet{nara12}, we derive surface densities for the gas mass for these systems, based on the CO($J=2\rightarrow1$) source sizes. Using metallicities of 1$\,$Z$_{\odot}$ and 0.02$\,$Z$_{\odot}$, the values range from $10^{4-5}$ M$_\odot$ pc$^{-2}$. When plotted in a Kennicutt-Schmidt diagram, the low metallicity assumption places the HyLIRGs closer to the trend at lower SFR, but all three assumptions mark the HyLIRGs as strong starbursts, i.e. well above main sequence galaxies. A possible limit on object metallicity was found by setting the flexible conversion factor to that of ULIRGs, yielding $Z\sim Z_{\odot}/3$.

A tentative east-west linear structure is seen in the CO($J=2\rightarrow1$) image of the SMG in BRI1202--0725, possibly a disk of size $\sim 0.8''$. This molecular disk is consistent with the east-west velocity gradient seen in the [CII] ALMA observations \citep{carn13}. 

For the QSO host galaxy in BRI1202--0725, the CO($J=2\rightarrow1$) emission is clearly
extended, although the signal-to-noise of the emission in the B--configuration 
provides only a lower  limit to the size of $0.5'' (\sim 3.3\,$kpc). 

For BRI1335--0417, we confirm the very extended CO emission to the north of the
QSO host galaxy, as seen in \citet{riec08}.  This extended
emission has a narrower velocity dispersion than for the main galaxy. 
The negligible velocity offset with respect
to the main galaxy and the lower velocity dispersion of this extended
gas suggests that it may be a remnant tidal feature from a previous
major merger of gas rich galaxies. This is based on radio observations, 
as past optical and submm observations lacked the necessary resolution to 
distinguish the northern extension. 

In all three sources, the extent of the 44$\,$GHz continuum emission
appears smaller than that of the CO($J=2\rightarrow1$) line emission,
or the area of active star formation is smaller than its fuel
supply. This suggests a varying level of excitation and star formation
activity across each source. The variation can be seen in
BRI1335--0417, which shows a southern core in 44$\,$GHz emission, but
no northern extension. Since both are seen in CO emission, this
suggests that the northern area is not starbursting. This size
discrepancy between low order CO and star formation has been noted for
other high redshift SFGs (star forming galaxies; e.g., as compiled by
\citealt{spil15}). Alternatively, the different sizes could simply be
a result of low signal-to-noise continuum observations not
detecting the lower optical depth outer sections of the star forming
regions.  Higher resolution, more sensitive ALMA observations of the
dust continuum emission are planned. These should answer this question
of the relative distributions of gas and star formation.

We also detect CO($J=2\rightarrow1$) emission from Ly$\alpha$-2 in the BRI1202--0725
system. The total gas mass is $(3.2\pm0.5)\times (\alpha_{CO}/0.8)\times10^{10}\,$M$_{\odot}$, and the gas depletion timescale is $2\times 10^8\times (\alpha_{CO}/0.8)\,$yr, 
where $\alpha_{CO}$ has units of M$_{\odot}$K$^{-1}$ km$^{-1}$ s pc$^{-2}$. Even assuming a low value of $\alpha_{CO}$ (compared to $\alpha_{CO}\sim4$ for normal galaxies; \citealt{bola13}), this galaxy
has a gas depletion timescale comparable to main sequence galaxies at
low and high redshift (e.g., $7\times10^8\,$years for $z=1-3$ MS SFGs, \citealt{tacc13}), and not the extremely short timescales
applicable to starbursts.

\citet{klam04} suggested that the extreme aspects of the BRI1202--0725 system might relate to star formation induced by a strong radio jet from the QSO.  While no such radio jet has yet been seen, the presence of  Ly$\alpha$-2 in the direction of the ouflow from the QSO seen in [CII] \citep{cari13}, is circumstantially suggestive of a hydrodynamic interaction as well as gravitational.

We have used our CO($J=2-1$) luminosities in conjunction with previous [CII] luminosities (\citealt{wagg12,wagg10,cari13}) and FIR luminosities (\citealt{carn13,cari02,will14}) to constrain the density and FUV radiation field of photodissociation regions (PDRs) in each source with the diagnostic plot of \citet{stac10}. Assuming thermalized ratios (L$_{CO(2-1)}$/L$_{CO(1-0)}$=4), this showed that PDRs in the three HyLIRGs showed similar densities and radiation fields to local ULIRGs and $z>2.3$ objects. When comparing the density and FUV level of PDRs in BRI1202-0725Ly$\alpha$-2 to other objects, an ambiguity arises. They are either similar to normal galaxies in level of FUV emission but are more dense or are similar to ULIRGs in density but feature weaker radiation fields. Current size constraints do not allow for concrete conclusions here.

Our data confirm that these HyLIRGs are extreme starbursts with short depletion timescales, as shown in the Kennicutt-Schmidt Relation. They are consistent with models that show SMG formation in short time periods \citep{nara15}, although they are at slightly higher redshift than expected. 

The data for these two HyLIRG systems suggest that they represent two
different stages in the evolution of extreme starbursts driven by
major gas rich mergers in the early Universe.  BRI1202--0725 appears to be a
relatively early stage merger, with a number of distinct galaxies
still observed in stars and gas.  There are clear indications of
strong gravitational interaction between the galaxies likely driving
the extreme starbursts, as well as possible evidence for a strong
QSO driven outflow assisting in quenching the star formation in the
QSO host. BRI1335--0417 appears to be a later stage merger, with just one
galaxy seen (the QSO host), plus what may be tidal remnants of the
merger seen in the extended cold gas.

The high luminosities, disrupted morphologies, evidence of gravitational interaction, and short gas depletion timescales of these objects suggest that they represent a transient, but highly star forming phase of early galaxy evolution. 

\vspace{2mm}

G. C. J. is grateful for support from NRAO through the Grote Reber Doctoral Fellowship Program. R. G. M.  acknowledges the UK Science and Technology Facilities Council (STFC). K. O. acknowledges the Kavli Institute Fellowship at the Kavli Institute for Cosmology in the University of Cambridge supported by the Kavli Foundation. The National Radio Astronomy Observatory is a facility of the National Science Foundation operated under cooperative agreement by Associated Universities, Inc. We thank all those involved in the VLA project for making these observations possible (project code 13A-012).

\appendix
\section{Additional Eddington Limits}
As another route to investigate if these sources are Eddington-limited, consider the basic Eddington luminosity criterion:
\begin{equation}
F_G=-F_{RAD}\rightarrow \frac{GMm}{r^2}=P_{RAD}A=\frac{L_{Edd}}{c}
\end{equation}
which can be reduced to $L_{Edd}=4\pi G M c \Sigma_{gas}$, in SI units. Switching the units of M, L$_{Edd}$, and $\Sigma_{Gas}$ to solar masses, solar luminosities, and solar masses per kpc$^2$, respectively, gives:
\begin{equation}
  L_{Edd}[L_{\odot}]=(2.71\times 10^{-6})M[M_{\odot}]\Sigma_{Gas}[M_{\odot}/kpc^2]
\end{equation}
(also given by equation 2 of \citealt{thom14}). Using the approximate gas or dynamical mass of each HyLIRG (10$^{11}\,$M$_{\odot}$) and our approximate gas densities (10$^{10-11}\,$M$_{\odot}$ kpc$^{-2}$) gives $L_{edd}\sim10^{15-16}\,$L$_{\odot}$. Since these sources have L$_{FIR}\sim10^{13}\,$L$_{\odot}$, they are radiating at $<1\%$ Eddington luminosity. Equation 20 of \citet{thom14}, which gives the fraction of Eddington luminosity as a function of SFR and gas surface densities, returns a similar percentage. The source Ly$\alpha$-2 is also sub-Eddington, using its size limits and estimate area densities.


\begin{thebibliography}{99}
\bibitem[Benford et al.(1999)]{benf99} 
Benford, D.~J., Cox, P., Omont, A., Phillips, T.~G., \& McMahon, R.~G.\ 1999, \apjl, 518, L65 

\bibitem[Blain et al.(2002)]{blai02}
Blain, A.~W., Smail, I., Ivison, R.~J., Kneib, J.-P., \& Frayer, D.~T.\ 2002, \physrep, 369, 111

\bibitem[Bolatto et al.(2013)]{bola13} 
Bolatto, A.~D., Wolfire, M., \& Leroy, A.~K.\ 2013, \araa, 51, 207 

\bibitem[Bothwell et al.(2010)]{both10}
Bothwell, M. S. et al. 2010, MNRAS, 405, 219

\bibitem[Capak et al.(2011)]{capa11} 
Capak, P.~L., Riechers, D., Scoville, N.~Z., et al.\ 2011, \nat, 470, 233 

\bibitem[Carilli et al.(1999)]{cari99}
Carilli, C. L., Menten, K. M., Yun, M. S. 1999, \apj, 521, L25

\bibitem[Carilli et al.(2002)]{cari02}
Carilli, C. L., Kohno, K., Kawabe, R., Ohta, K., Henkel, C., et al. 2002, \aj, 123, 1838

\bibitem[Carilli et al.(2013)]{cari13}
Carilli, C. L., Riechers, D., Walter, F., Maiolino, R., Wagg, J., et al. 2013, \apj, 763, 120

\bibitem[Carilli \& Walter(2013)]{cariw13}
Carilli, C.~L. \& Walter, F.\ 2013, \araa, 51, 105

\bibitem[Carniani et al.(2013)]{carn13}
Carniani, S., Marconi, A., Biggs, A., Cresci, G., Cupani, G., et al. 2013, A\&A, 559, A29

\bibitem[Casey et al.(2012)]{case12} 
Casey, C.~M., Berta, S., B{\'e}thermin, M., et al.\ 2012, \apj, 761, 140

\bibitem[Casey et al.(2014)]{case14}
Casey, C.~M., Narayanan, D., \& Cooray, A.\ 2014, \physrep, 541, 45

\bibitem[Chapman et al.(2009)]{chap09} 
Chapman, S.~C., Blain, A., Ibata, R., et al.\ 2009, \apj, 691, 560 

\bibitem[Daddi et al.(2009)]{dadd09} 
Daddi, E., Dannerbauer, H., Stern, D., et al.\ 2009, \apj, 694, 1517 

\bibitem[Decarli et al.(2014)]{deca14}
Decarli, R., Walter, F., Carilli, C., et al.\ 2014, \apjl, 782, L17

\bibitem[Engel et al.(2010)]{enge10} 
Engel, H., Tacconi, L.~J., Davies, R.~I., et al.\ 2010, \apj, 724, 233 

\bibitem[Fontana et al.(1996)]{font96} 
Fontana, A., Cristiani, S., D'Odorico, S., Giallongo, E., \& Savaglio, S.\ 1996, \mnras, 279, L27 

\bibitem[Fontana et al.(1998)]{font98} 
Fontana, A., D'Odorico, S., Giallongo, E., et al.\ 1998, \aj, 115, 1225

\bibitem[Freundlich et al.(2013)]{freu13}
Freundlich et al. 2013, A\&A, 553, A130

\bibitem[Genzel et al.(2010)]{genz10}
Genzel, R. et al. 2010, MNRAS, 407, 2091

\bibitem[Greve et al.(2005)]{grev05} 
Greve, T.~R., Bertoldi, F., Smail, I., et al.\ 2005, \mnras, 359, 1165 

\bibitem[Guilloteau et al.(1997)]{guil97}
Guilloteau, S., Omont, A., McMahon, R. G., Cox, P., Petitjean, P. 1997, A\&A, 328, L1

\bibitem[Hodge et al.(2012)]{hodg12} 
Hodge, J.~A., Carilli, C.~L., Walter, F., et al.\ 2012, \apj, 760, 11

\bibitem[Hodge et al.(2015)]{hodg15}
Hodge, J. A., Riechers, D., Decarli, R., Walter, F., Carilli, C. L., et al. 2015, \apjl, 798, L18

\bibitem[Hu et al.(1996)]{hu96} 
Hu, E.~M., McMahon, R.~G., \& Egami, E.\ 1996, \apjl, 459, L53 

\bibitem[Hu et al.(1997)]{hu97}
Hu, E. M., McMahon, R. G., \& Egami, E. 1997, in The Hubble Space Telescope
and the High-Redshift Universe, eds. N
. R. Tanvir, A. Aragon-Salamanca, \&
J. V. Wall (Singapore: World Scientific), 91

\bibitem[Iono et al.(2006)]{iono06}
Iono, D., Yun, M.~S., Elvis, M., et al.\ 2006, \apjl, 645, L97 

\bibitem[Irwin et al.(1991)]{irwi91} 
Irwin, M., McMahon, R.~G., \& Hazard, C.\ 1991, The Space Distribution of QSOs, 21, 117

\bibitem[Isaak et al.(1994)]{isaa94} 
Isaak, K.~G., McMahon, R.~G., Hills, R.~E., \& Withington, S.\ 1994, \mnras, 269, L28

\bibitem[Ivison et al.(2012)]{ivis12}
Ivison, R.~J., Smail, I., Amblard, A., et al.\ 2012, \mnras, 425, 1320 

\bibitem[Karim et al.(2016, in preparation)]{kari16}
Karim, A. et al.\ 2016 \textit{in prep}

\bibitem[Kawabe et al.(1999)]{kawa99} 
Kawabe, R., Kohno, K., Ohta, K., \& Carilli, C.\ 1999, Highly Redshifted Radio Lines, 156, 45

\bibitem[Kennicutt(1998)]{kenn98}
Kennicutt, R. C. 1998, ApJ, 498, 541

\bibitem[Kennicutt et al.(2007)]{kenn07}
Kennicutt, R. C. et al. 2007, ApJ, 671, 333

\bibitem[Kennicutt \& Evans(2012)]{kenn12}
Kennicutt, R. C. \& Evans, N. J. 2012, ARA\&A, 50, 531
  
\bibitem[Kimball et al.(2015)]{kimb15}
Kimball, A.~E., Lacy, M., Lonsdale, C.~J., \& Macquart, J.-P.\ 2015, \mnras, 452, 88

\bibitem[Klamer et al.(2004)]{klam04}
Klamer, I.~J., Ekers, R.~D., Sadler, E.~M., \& Hunstead, R.~W.\ 2004, \apjl, 612, L97 

\bibitem[Kormendy \& Ho(2013)]{korm13}
Kormendy, J., \& Ho, L.~C.\ 2013, \araa, 51, 511 

\bibitem[Leech et al.(2001)]{leec01}
Leech, K.~J., Metcalfe, L., \& Altieri, B.\ 2001, \mnras, 328, 1125 

\bibitem[Lequeux et al.(1979)]{lequ79}
Lequeux, J., Peimbert, M., Rayo, J.~F., Serrano, A. \& Torres-Peimbert, S.\ 1979, \aap, 80, 155

\bibitem[McMahon et al.(1994)]{mcma94}
McMahon, R.~G., Omont, A., Bergeron, J., Kreysa, E., \& Haslam, C.~G.~T.\ 1994, \mnras, 267, L9 

\bibitem[Miley \& De Breuck(2008)]{mile08}
Miley, G., \& De Breuck, C.\ 2008, \aapr, 15, 67 

\bibitem[Momjian et al.(2005)]{momj05}
Momjian, E., Carilli, C. L., Petric, A. 2005 AJ, 129, 1809

\bibitem[Momjian et al.(2007)]{momj07}
Momjian, E., Carilli, C. L., Riechers, D. A., Walter, F. 2007 \apj, 134, 694

\bibitem[Narayanan et al.(2012)]{nara12} 
Narayanan, D., Krumholz, M.~R., Ostriker, E.~C., \& Hernquist, L.\ 2012, \mnras, 421, 3127 

\bibitem[Narayanan et al.(2015)]{nara15} 
Narayanan, D., Turk, M., Feldmann, R., et al.\ 2015, \nat, 525, 496

\bibitem[Ohta et al.(1996)]{ohta96}
Ohta, K., Yamada, T., Nakanishi, K., Kohno, K., Akiyama, M., Kawabe, R. 1996, \apjl, 382, 426

\bibitem[Ohta et al.(2000)]{ohta00}
Ohta, K., Matsumoto, T., Maihara, T., et al.\ 2000, \pasj, 52, 557

\bibitem[Omont et al.(1996a)]{omonA96} 
Omont, A., McMahon, R.~G., Cox, P., et al.\ 1996, \aap, 315, 1

\bibitem[Omont et al.(1996b)]{omonN96}
Omont A., Petitjean P., Guilloteau S., McMahon, R. G., Solomon, P. M., P\'econtal, E. 1996, Nat 382, 428

\bibitem[Perley \& Butler(2013)]{perl13} 
Perley, R.~A., \& Butler, B.~J.\ 2013, \apjs, 204, 19 

\bibitem[Petitjean et al.(1996)]{peti96} 
Petitjean, P., P{\'e}contal, E., Valls-Gabaud, D., \& Chariot, S.\ 1996, \nat, 380, 411

\bibitem[Planck Collaboration et al.(2015)]{plan15} 
Planck Collaboration, Ade, P.~A.~R., Aghanim, N., et al.\ 2015, arXiv:1502.01589 

\bibitem[Renzini(2006)]{renz06}
Renzini, A. 2006, ARA\&A, 44, 141

\bibitem[Riechers et al.(2006)]{riec06}
Riechers, D. A., Walter, F., Carilli, C. L., Knudsen, K. K., Lo, K. Y. 2006, \apj, 650, 604

\bibitem[Riechers et al.(2008)]{riec08}
Riechers, D. A., Walter, F., Carilli, C. L., Bertoldi, F., Momjian, E. 2008, \apj, 686, L9

\bibitem[Riechers et al.(2011)a]{riec11L31}
Riechers, D.~A., Hodge, J., Walter, F., Carilli, C.~L., \& Bertoldi, F.\ 2011a, \apjl, 739, L31

\bibitem[Riechers et al.(2011)b]{riec11L32}
Riechers, D.~A., Carilli, C.~L., Maddalena, R.~J., et al.\ 2011b, \apjl, 739, L32 

\bibitem[Salom{\'e} et al.(2012)]{salo12} 
Salom{\'e}, P., Gu{\'e}lin, M., Downes, D., et al.\ 2012, \aap, 545, A57

\bibitem[Sanders \& Mirabel(1996)]{sand96} 
Sanders, D.~B., \& Mirabel, I.~F.\ 1996, \araa, 34, 749 

\bibitem[Scoville(2012)]{scov12}
Scoville, N. 2012, arXiv:1210.6990

\bibitem[Shapley(2011)]{shap11}
Shapley, A.~E.\ 2011, \araa, 49, 525

\bibitem[Shields et al.(2006)]{shie06} 
Shields, G.~A., Menezes, K.~L., Massart, C.~A., \& Vanden Bout, P.\ 2006, \apj, 641, 683

\bibitem[Spilker et al.(2015)]{spil15}
Spilker, J.~S., Aravena, M., et al.\ 2015, \apj, 811, 124

\bibitem[Stacey et al.(2010)]{stac10}
Stacey, G.~J., Hailey-Dunsheath, S., et al.\ 2010, \apj, 724, 957

\bibitem[Tacconi et al.(2008)]{tacc08}
Tacconi, L.~J., Genzel, R., Smail, I., et al.\ 2008, \apj, 680, 246 

\bibitem[Tacconi et al.(2013)]{tacc13}
Tacconi, L.~J., Neri, R., Genzel, R., et al.\ 2013, \apj, 768, 74 

\bibitem[Thompson et al.(2005)]{thom05} 
Thompson, T.~A., Quataert, E., \& Murray, N.\ 2005, \apj, 630, 167 

\bibitem[Thompson \& Krumholz(2014)]{thom14}
Thompson, T.~A., \& Krumholz, M.~R.\ 2014, arXiv:1411.1769

\bibitem[van den Bosch et al.(2012)]{vand12}
van den Bosch, R.~C.~E., Gebhardt, K., G{\"u}ltekin, K., et al.\ 2012, \nat, 491, 729

\bibitem[Wagg et al.(2010)]{wagg10}
Wagg, J., Carilli, C. L., Wilner, D. J., Cox, P., De Breuck, C. et al. 2010, A\&A, 519, L1

\bibitem[Wagg et al.(2012)]{wagg12}
Wagg, J., Wiklind, T., Carilli, C., Espada, D., Peck, A., et al. 2012, \apj, 752, L30

\bibitem[Wagg et al.(2014)]{wagg14}
Wagg, J., Carilli, C. L., Aravena, M., et al. 2014, \apj, 783, 71

\bibitem[Walter et al.(2004)]{walt04} 
Walter, F., Carilli, C., Bertoldi, F., et al.\ 2004, \apjl, 615, L17 

\bibitem[Wang et al.(2013)]{wang13}
Wang, R., Wagg, J., Carilli, C.~L., et al.\ 2013, \apj, 773, 44 

\bibitem[Williams et al.(2014)]{will14} 
Williams, R.~J., Wagg, J., Maiolino, R., et al.\ 2014, \mnras, 439, 2096

\bibitem[Willott et al.(2015)]{will15}
Willott, C.~J., Carilli, C.~L., Wagg, J., \& Wang, R.\ 2015, \apj, 807, 180

\bibitem[Yun et al.(1999)]{yun99} 
Yun, M.~S., Scoville, N.~Z., \& Evans, A.~S.\ 1999, Highly Redshifted Radio Lines, 156, 58 

\bibitem[Yun et al.(2000)]{yun00}
Yun, M.~S., Carilli, C.~L., Kawabe, R., et al.\ 2000, \apj, 528, 171 

\end{thebibliography}
\end{document}